\documentclass[aps,prb,amsmath,amssymb,bibnotes,reprint,longbibliography,superscriptaddress]{revtex4-1}

\usepackage{t1enc}
\usepackage[utf8]{inputenc}
\usepackage{amsmath}
\usepackage{bbold}

\usepackage{graphicx}
\usepackage{siunitx}
\usepackage{hyperref}
\usepackage{braket}
\usepackage{hhline}

\renewcommand\vec{\mathbf}

\usepackage{xcolor}

\begin{document}

\title{Braiding-based quantum control
of a Majorana qubit built from quantum dots}

\author{P\'{e}ter Boross}
\affiliation{Department of Theoretical Physics, Institute of Physics, Budapest University of Technology and Economics, M\H{u}egyetem rkp. 3., H-1111 Budapest, Hungary}
\affiliation{Institute for Solid State Physics and Optics, Wigner Research Centre for Physics, H-1525 Budapest P.O. Box 49, Hungary}

\author{Andr\'as P\'alyi}
\affiliation{Department of Theoretical Physics, Institute of Physics, Budapest University of Technology and Economics, M\H{u}egyetem rkp. 3., H-1111 Budapest, Hungary}
\affiliation{MTA-BME Quantum Dynamics and Correlations Research Group, M\H{u}egyetem rkp. 3., H-1111 Budapest, Hungary}

\date{\today}

\begin{abstract}
Topology-related ideas might lead to noise-resilient
quantum computing. 
For example, it is expected that the slow spatial exchange 
(`braiding') of Majorana zero modes in  superconductors
yields quantum gates that are robust against
disorder. 
Here, we report our numerical experiments, which describe the dynamics of a Majorana qubit built from quantum dots controlled by time-dependent gate voltages. 
Our protocol incorporates 
non-protected control,
braiding-based protected control, and readout,
of the Majorana qubit. 
We use the Kitaev chain 
model for the simulations, and 
focus on the case when the main source of errors is
quasistatic charge noise affecting the
hybridization energy splitting of the Majorana modes.
We provide quantitative guidelines to suppress both diabatic errors and disorder-induced qubit dephasing, such that a fidelity plateau is observed as the hallmark of the topological quantum gate.
Our simulations predict realistic features
that are expected to be seen in future braiding 
experiments with Majorana zero modes
and other topological qubit architectures.
\end{abstract}

\maketitle

\tableofcontents

\section{Introduction}

Majorana Zero Modes (MZMs) as bound states in topological superconductors might be used as building blocks of future quantum technology, enabling topologically protected quantum-logical qubit operations \cite{IvanovPRL2001,KitaevPU2001,AliceaNatPhys2011}.
This opportunity has triggered intense research efforts in the past decade\cite{MourikScience2012,AlbrechtNat2016,AliceaRPP2012,BeenakkerRMP2015,DasSarmaNPJQI2015,BeenakkerNatPhys2016,AguadoReview,LutchynNatRevMat2018,BeenakkerSciPost2020,PradaReview}. 
The protection of the quantum information encoded in such a Majorana qubit is, however, incomplete. \cite{ScheurerPRB2013,AseevPRB2018,AseevPRB2019,WalterPRB2012,KnappPRB2018,AasenPRX2016,KarzigPRB2017,HellPRB2016,RahmaniPRB2017,Breckwoldt_2022,BauerSciPost2018,FulgaPRB2013,ZhangPRA2019,KornichPRL2021}
To assess the technological potential of Majorana qubits, it is critical to understand their decoherence mechanisms. 

Due to their topological protection, Majorana qubits might serve as long-lived quantum memory elements, hence the decoherence properties of idle qubits is of interest \cite{BorossPRB2022,AseevPRB2018,AseevPRB2019,AasenPRX2016,BrouwerPRL2011,GoldsteinPRB2011,SchmidtPRB2012,BudichPRB2012,RainisPRB2012,PedrocchiPRL2015,KnappPRB2018,BauerSciPost2018,LaiPRB2018,MishmashPRB2020}.
Majorana qubits could also serve as building blocks of quantum processors, where they are subject to electromagnetic control fields, e.g., controlling the spatial exchange of the MZMs\cite{AliceaNatPhys2011,TutschkuPRB2020}.
Such a MZM exchange can yield a single-qubit $\pi/2$ gate, and can also be useful to perform two-qubit gates, e.g., a CNOT \cite{ZilberbergPRA2008,TutschkuPRB2020,BeenakkerPRL2004}.
(Note that other works propose `braiding without braiding', i.e,. measurement-based quantum gates that avoid spatial exchange MZMs\cite{VijayPRB2016,KarzigPRB2017,PluggeNJP2017}.)

Reference \onlinecite{AliceaNatPhys2011} proposed to braid MZMs in one-dimensional (1D) topological superconductors to achieve topologically protected quantum gates. 
Advantages and limitations of this topological protection of braiding-based gates has been the subject of many theoretical works since then.
Models of MZM braiding range from the effective description restricted to the degenerate subspace\cite{FulgaPRB2013,KarzigPRB2015,KnappPRX2016,RahmaniPRB2017,ZhangPRA2019,NagPRB2019,StengerPRR2021}, through the Kitaev chain model \cite{AmorimPRB2015,ChengPLA2016,SekaniaPRB2017,Breckwoldt_2022}, to the Rashba wire model \cite{HarperPRR2019,TutschkuPRB2020}.

MZMs are also predicted in engineered topological superconductors, where an effective Kitaev chain is formed by a register of quantum dots proximitized by nearby superconductors.\cite{Leijnse_2012,SauNatComm2012,Fulga_2013,TsintzisPRB2022,ChunXiaoLiuPRL2022}
Recent experimental progress with quantum-dot chains, featuring charge shuttling in a 9-dot array \cite{MillsNatComm2019},
spin qubit operation in 6-dot arrays \cite{PhilipsArxiv2022,Weinstein},
the triple Andreev dot chain \cite{WuArxiv2021},
and the realization of the mininal Kitaev chain \cite{Dvir}
strengthen the feasibility of the quantum-dot approach to MZMs.
In such quantum dots, often used as spin qubit registers, a key qubit deocoherence mechanism is the fluctuation of the on-site energies of the dots, attributed to electromagnetic fluctuations of the environment, including $1/f$ charge noise \cite{MishmashPRB2020,CywinskiPRB2008,DialPRL2013,YonedaNatNano2018,ShnirmanPhysScri2002,FreemanAPL2016,HetenyiPRB2019,CywinskiPRB2020}. 

In this work, we propose a few-dot setup and a control scheme, which could be used to experimentally demonstrate braiding-based gates with MZMs based on quantum dot arrays \cite{Leijnse_2012,SauNatComm2012,Fulga_2013}.
We perform numerical experiments to predict the quality of braiding-based gates in the presence of charge noise, which we incorporate in our model as quasistatic disorder.
Our numerical results show diabatic errors for short braiding times as well as qubit dephasing effects for long braiding times. 
We identify an experimentally relevant parameter range where a future experiment can find `fidelity plateaus' as fingerprints of topologically protected quantum gates.

The rest of the paper is organized as follows.
In Sec.~\ref{sec:model}, we introduce the setup of our numerical experiment and highlight the main steps of our braiding-based protocol. In Sec.~\ref{sec:clean}, we describe the non-adiabatic dynamics of the system in the absence of any disturbances. In Sec.~\ref{sec:disorder}, we show the effect of the quasistatic disorder on our protocol. We discuss implications and follow-up ideas in Sec.~\ref{sec:discussion}, and conclude in Sec.~\ref{sec:conclusions}.

% = = = = 

\section{Protocol to demonstrate a braiding-based $\pi/2$ gate of a Majorana qubit}
\label{sec:model}

\begin{figure*}[t]
	\centering
	\includegraphics[width=2\columnwidth]{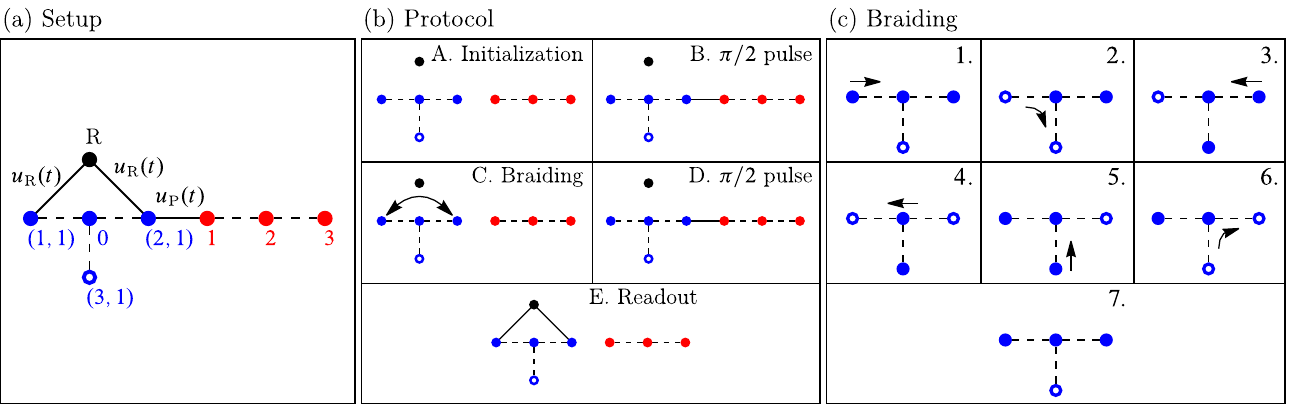}
	\caption{
	Protocol to demonstrate a braiding-based $\pi/2$ gate 
	on a Majorana qubit.
	(a) Blue: Y junction built from Kitaev chains.
    Red: a straight Kitaev wire. Blue and red together form a Majorana qubit. Black: readout dot. 
    (b) The 5-step protocol, analogous to a Ramsey 
    protocol, aiming to detect the braiding-based 
    $\pi/2$ gate: 
    (A) Initialization,
    (B) First $\pi/2$ pulse, 
    (C) Braiding,
    (D) Second $\pi/2$ pulse,
    (E) Readout.
    (c) Procedure of a single MZM exchange used in 
    step (C) Braiding. 
	}
	\label{fig:setup}
	\centering
\end{figure*}

Here we introduce a model and a protocol, suitable for the experimental demonstration of the braiding-based $\pi/2$ quantum gate of a Majorana qubit. 
The setup is sketched in Fig.~\ref{fig:setup}a. It consists of 3 units: a Y junction built from Kitaev chains (blue), a straight Kitaev wire (red), and a readout dot (black). 
Dashed lines denote connections via electron tunneling and Cooper-pair creation and annihilation. Solid lines denote connections via tunneling only. Filled and empty circles depict different on-site energies.
This setup consists of two Kitaev chains (blue filled circles and red filled circules), and hence it can host a Majorana qubit in its ground-state subspace.  
Based on Refs.~\onlinecite{Leijnse_2012,SauNatComm2012,Fulga_2013,TsintzisPRB2022,Dvir}, we envision that proximitized quantum dot arrays can realize such a few-site Kitaev model.

The Y junction consists of three regular Kitaev chains (legs) and a central site. Each Kitaev chain can be described by the following Hamiltonian\cite{KitaevPU2001}:
\begin{align}
\label{eq:Kitaevwire}
	H_c^{(n,N_\text{c})} =& \sum_{i=1}^{N_\text{c}-1}{\left(v c^\dag_{n,i} c^{\vphantom{\dag}}_{n,i+1}+\Delta_n c^\dag_{n,i} c^\dag_{n,i+1}+\text{h.c.} \right)} \nonumber \\
	& + \sum_{i=1}^{N_\text{c}}{\mu_{n,i}(t) c^\dag_{n,i} c^{\vphantom{\dag}}_{n,i}},
\end{align}
where $N_\text{c}$ is the length of the legs (it is set to $N_\text{c} = 1$ in Fig.~\ref{fig:setup}), $c^{\dag}_{n,i}$ and $c^{\vphantom{\dag}}_{n,i}$ are the creation and annihilation operators on the $i$th site of the $n$th chain ($n \in \{1,2,3\}$), $v$ denotes the nearest-neighbor hopping amplitude, $\Delta_n$ is the $p$-wave superconducting pairing amplitude in the $n$th chain, and $\mu_{n,i}(t)$ is the site- and time-dependent on-site energy of the $i$th site of the $n$th chain. 

Based on the Hamiltonians of the legs in Eq.~\eqref{eq:Kitaevwire}, the Hamiltonian of the Y junction is written as 
\begin{align}
	H_Y =& \sum_{n=1}^{3}{H_c^{(n,N_\text{c})}} + \mu_{0} c^\dag_{0} c^{\vphantom{\dag}}_{0} \nonumber \\
	&+ \sum_{n=1}^{3}{\left( v c^\dag_{0} c^{\vphantom{\dag}}_{n,1} + \Delta_n c^\dag_{0} c^{\dag}_{n,1} + \text{h.c.}\right)},
\end{align}
where index $0$ denotes the central site, and the superconducting pair potential is $\Delta_1 = \Delta e^{i \varphi}$, $\Delta_2 = \Delta e^{-i \varphi}$ and $\Delta_3 = -\Delta$, with $\Delta > 0$.
For concreteness, we set $\varphi = \pi/2$ in our simulations.

Another building block of the Majorana qubit is a straight Kitaev wire, depicted as filled red circles in Fig.~\ref{fig:setup}. Its length is chosen, for simplicity, to have the same length as the topological region of the Y-junction, which is 3 in the case of $N_\text{c} = 1$. Thus the corresponding Hamiltonian is 
\begin{align}
\label{eq:hc}
	H_W =& \sum_{i=1}^{2N_\text{c}}{\left(v c^\dag_{i} c^{\vphantom{\dag}}_{i+1}+\Delta_W c^\dag_{i} c^\dag_{i+1}+\text{h.c.} \right)} \nonumber \\
	& + \sum_{i=1}^{2N_\text{c}+1}{\mu_{i} c^\dag_{i} c^{\vphantom{\dag}}_{i}},
\end{align}
where $\Delta_W=\Delta e^{-i \varphi}$.

With the purpose of reading out the parity of the Y junction, an additional site, the \emph{readout dot} \cite{SzechenyiPRB2020,GharaviPRB2016,AasenPRX2016,KarzigPRB2017} denoted by `R' in Fig.~\ref{fig:setup}a, is coupled to the system. The Hamiltonian of the full system reads:
\begin{align}
    \label{eq:fullFockHamiltonian}
	H =& H_Y + H_W + \mu_R c^\dag_{R} c^{\vphantom{\dag}}_{R} + \left( u_\text{P}(t) c^\dag_{2,N_\text{c}} c^{\vphantom{\dag}}_{1} + \text{h.c.}\right) \nonumber \\
	&+ \sum_{n=1}^{2}{\left( u_\text{R}(t) c^\dag_R c^{\vphantom{\dag}}_{n,N} + \text{h.c.}\right)}.
\end{align}
Here, $u_\text{P}(t)$ is the hopping amplitude between the rightmost site of the Y junction and the leftmost site of the straight Kitaev wire, which is required in the (non-protected) $\pi/2$ pulses of the experimental protocol, see below. 
Furthermore, $u_\text{R}(t)$ is the tunneling amplitude between the readout dot and the two ends of the $Y$-junction, which is utilized for parity-to-charge conversion in the readout step of the experimental protocol. 

The Y junction and the straight Kitaev wire can host two MZMs each, and their composite system can host a Majorana qubit. By default, both the Y junction and the straight Kitaev wire are tuned to their topological fully dimerized limit, and hence the ground state of their composite system is fourfold degenerate. The four ground states will be denoted as $\ket{\text{e}, \text{e}}$, $\ket{\text{e}, \text{o}}$, $\ket{\text{o}, \text{e}}$, $\ket{\text{o}, \text{o}}$, with $\text{e}$ ($\text{o}$) being a reference to the even (odd) fermion-number parity of each unit. The topological fully dimerized limit is defined by setting 
(i) the hopping amplitudes and the absolute value of the superconducting pair potentials equal to each other, i.e.~$\Delta=v$, and
(ii) the on-site potentials to zero.
This limit implies MZMs that are perfectly localized at the end sites of the topological regions.

\begin{figure*}[t]
	\centering
	\includegraphics[width=1.5\columnwidth]{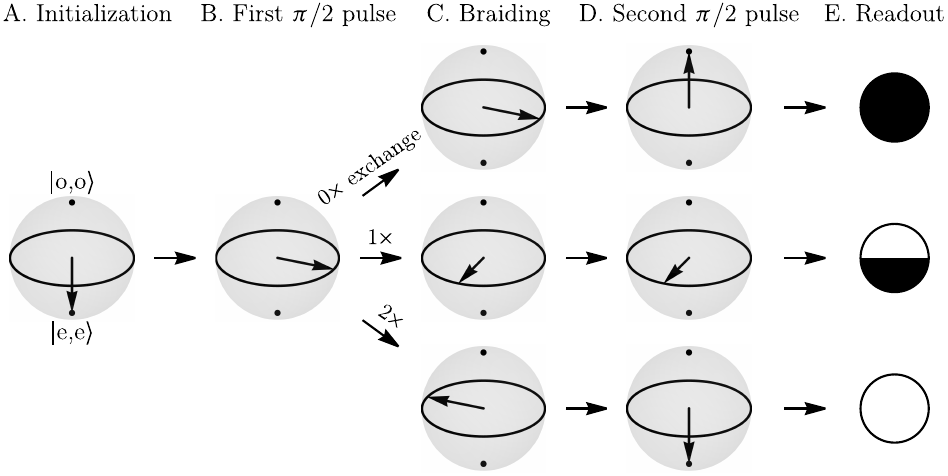}
	\caption{Ramsey-type experiment to detect the braiding-based $\pi/2$ gate on the Majorana qubit. 
	Error-free evolution of the qubit Bloch vector, shown after each step (A)-(D) of the protocol, cf. Table \ref{tab:idealprotocol}.
	Qubit basis states are $\ket{\text{e,e}}$ and $\ket{\text{o,o}}$.
	Last column shows the expected value of the readout dot occupation at the and of the protocol. The three paths correspond to the protocol with 0, 1 and 2 exchanges.}
	\label{fig:protocol}
	\centering
\end{figure*}

\begin{table*}[t]
\centering
\setlength\extrarowheight{5pt}
\begin{tabular}{p{3.5cm}|c|c|c|}
\cline{2-4}
 &
  $0\times$ exchange &
  $1\times$ exchange &
  $2\times$ exchanges \\[5pt] \hhline{-|===|}
\multicolumn{1}{|c||}{A. Initialization} &
  \multicolumn{3}{c|}{$\ket{\text{e},\text{e},0}$} \\[5pt] \hline
\multicolumn{1}{|c||}{B. First $\pi/2$ pulse} &
  \multicolumn{3}{c|}{$\frac{1}{\sqrt{2}}\ket{\text{e},\text{e},0}+\frac{1}{\sqrt{2}}\ket{\text{o},\text{o},0}$} \\[5pt] \hline
\multicolumn{1}{|c||}{C. Braiding} &
  $\frac{1}{\sqrt{2}}\ket{\text{e},\text{e},0}+\frac{1}{\sqrt{2}}\ket{\text{o},\text{o},0}$ &
  $\frac{1}{\sqrt{2}}\ket{\text{e},\text{e},0}+\frac{i}{\sqrt{2}}\ket{\text{o},\text{o},0}$ &
  $\frac{1}{\sqrt{2}}\ket{\text{e},\text{e},0}-\frac{1}{\sqrt{2}}\ket{\text{o},\text{o},0}$ \\[5pt] \hline
\multicolumn{1}{|c||}{D. Second $\pi/2$ pulse} &
  $\ket{\text{o},\text{o},0}$ &
  $\frac{1-i}{2}\ket{\text{e},\text{e},0}+\frac{1+i}{2}\ket{\text{o},\text{o},0}$ &
  $\ket{\text{e},\text{e},0}$ \\[5pt] \hline
\multicolumn{1}{|c||}{E. Readout} &
  $i\ket{\text{e},\text{o},1}$ &
  $\frac{1-i}{2}\ket{\text{e},\text{e},0}-\frac{1-i}{2}\ket{\text{e},\text{o},1}$ &
  $\ket{\text{e},\text{e},0}$ \\[5pt] \hline
\end{tabular}
\caption{Error-free evolution of
the many-body wave function during the Ramsey-type protocol. Three columns correspond to the protocol with 0, 1 and 2 exchanges. The wave function shown in each row is the state at the end of that step. \label{tab:idealprotocol}}
\end{table*}

The protocol we propose to demonstrate braiding is similar to the well-known Ramsey experiment, as shown in Fig.~\ref{fig:protocol} and Table \ref{tab:idealprotocol}. 
The key difference is that most often the Ramsey experiment aims to characterize unwanted dephasing dynamics (see, e.g., Fig.~10 of Ref.~\onlinecite{AasenPRX2016}), whereas here we use this scheme to characterise an intentional, braiding-based $\pi/2$ gate.

Our Ramsey-type protocol consists of five steps: (A) initialization, (B) first $\pi/2$ pulse, (C) braiding, (D) second $\pi/2$ pulse, and (E) readout, detailed in the subsections below.
The development of the many-body wave function at the key points of the protocol is shown in Table \ref{tab:idealprotocol}, and the development of the corresponding Majorana qubit polarization vector is shown in Fig.~\ref{fig:protocol}.
As seen in Table \ref{tab:idealprotocol} and Fig.~\ref{fig:protocol}, we consider three different cases: when no braiding is done ($0 \times$ exchange), 
when a single exchange is performed ($1 \times$ exchange), 
and when two exchanges are performed ($2 \times$ exchange).
In the Table, we use the notation $\ket{p_\text{Y},p_\text{W},n_\text{R}}$, where $p_\text{Y} \in \{\text{e}, \text{o}\}$ ($p_\text{W} \in \{\text{e}, \text{o}\}$) is the parity of Y junction (straight wire), and $n_\text{R} \in \{0,1\}$ is the occupation of the readout dot. 
Table \ref{tab:idealprotocol} shows the wave functions in an idealised, error-free case, when there is no disorder, no timing error, no leakage from the computational subspace, etc. 
Fig.~\ref{fig:protocol} shows the polarization vector of the Majorana qubit after each of the first four steps of the protocol, with the last column showing the readout dot occupation expectation value, which can be measured upon readout. 

The last step of the protocol is the measurement of the charge of the readout dot.
The measurement probabilities can be read off the last row (`Readout') of Table \ref{tab:idealprotocol} as follows. 
For $0\times$ exchange, the measurement probability of finding a charge in R is $1$, 
for $1\times$ exchange it is $0.5$, and for 
$2 \times$ exchange it is $0$.
This probability (which is the same as the charge expectation value of the readout dot) carries the parity information of the Y junction: if it is zero, that signals that the Y junction was in the even state after the second $\pi/2$ pulse, and before starting the parity-to-charge conversion (e.g., $2\times$ exchange); if it is one, that signals the odd parity (e.g., $0\times$ exchange).

In what follows, we describe the 5 steps of the protocol in detail.

\subsection{Initialization}

At the beginning of the protocol, the coupling between the Y-junction and the straight wire, and the coupling between the Y-junction and the readout dot, are turned off, i.e. $u_\text{P}=0$ and $u_\text{R}=0$. All the on-site potentials are set to zero, except in the third leg of the Y-junction, where they are set to the value $\xi=4v$, adjusted well over the critical value $2v$.
We assume that the Majorana qubit is initialized in the state $\ket{\text{e}, \text{e}}$. 
This implies that the physically available part of the 4-dimensional ground-state subspace is the 2-dimensional subspace spanned by the globally even ground states $\ket{\text{e}, \text{e}}$ and $\ket{\text{o}, \text{o}}$.
The two states form the computational basis for the Majorana qubit.
We also assume that the dot is initialized to be empty.

\subsection{First $\pi/2$ pulse}

The second step is a non-protected $\pi/2$ rotation
of the Majorana qubit, i.e., a rotation in the subspace spanned by $\ket{\text{e}, \text{e}}$ and $\ket{\text{o}, \text{o}}$.
We will refer to this as a rotation around the $y$ axis of the Bloch sphere.
(Note that this is an implicit condition for the relative global phase of the two qubit basis states, which we have not defined explicitly.)
Here, a balanced superposition of the basis states is achieved by switching on the tunnel coupling $u_\text{P}$. We apply a sine-squared shaped pulse which has the form 
\begin{equation}
    u_\text{P}(t) = u_\text{P,max} \sin^2(\pi t/T_\text{P}),\quad\text{if }0\leq t\leq T_\text{P},
\end{equation}
where duration $T_\text{P}=h/(4u_\text{P,max})$ of the pulse is set to provide a $\pi/2$ gate. To avoid quasiparticle excitation, we use a weak pulse, $u_\text{P,max}=0.25v$.
Note that this gate is not topologically protected\cite{FlensbergPRL2011,TommyLiPRB2018,BauerSciPost2018}; i.e., errors in the tunnel pulse strength or duration lead to gate errors that are not suppressed by increasing the system size or slower operation.

\subsection{Braiding}
The next step is the braiding of the MZMs localized on site $(1,1)$ and $(2,1)$, making use of the site $(3,1)$ of the Y junction. MZMs can be exchanged by means of the steps shown in Fig.~\ref{fig:setup}(c). 
Moving MZMs is realized by ramping up (down) the on-site energies to the value $\xi=4v$ (zero). These we do by using sine-squared pulses. The shape of the pulse for the ramp-up reads
\begin{equation}
    \mu_{n,i}(t) = \xi \sin^2 \left( \frac{\pi t}{2 T_\text{ramp}}\right), \quad\text{if }0\leq t\leq T_\text{ramp},
\end{equation}
where $T_\text{ramp}$ is the ramping time, i.e., the duration through which an onsite potential is varied. 
For longer chains, i.e.~$N_\text{c}>1$, ramp-up and ramp-down of on-site potentials in each leg is performed consecutively, site-by-site. Adiabatic exchange of the modes will create a $\pi/2$ phase difference \cite{AliceaNatPhys2011} between the basis state $\ket{\text{e}, \text{e}}$ and $\ket{\text{o}, \text{o}}$.
This phase difference should be robust against imperfections of the path of control parameters, including quasistatic disorder; characterising this robustness is one of the goals of the numerical experiments in the next sections.  

As we argue below, performing this protocol with two MZM exchange, providing a $\pi$ gate, should be used as an important control experiment. We will denote the number of exchanges by $N_\text{E}$. 
Furthermore, as seen from Fig.~\ref{fig:setup}c, the full duration of the Braiding step is $T_\text{B}=6N_\text{c}N_\text{E}T_\text{ramp}$.
Note also that we use the term `exchange' to denote a single (clockwise or counterclockwise) exchange of two MZMs, whereas we use the term `braiding' to describe any combination of MZM exchanges.

\subsection{Second $\pi/2$ pulse}

After the braiding step, a second non-protected $\pi/2$ pulse is applied, to do a rotation around the y axis. 
Naturally, the many-body state after this step (row 4 of Table \ref{tab:idealprotocol}) does depend on the number of exchanges done during Braiding.

\subsection{Readout}

In our protocol, readout of the parity of the Y junction is based on parity-to-charge conversion \cite{SzechenyiPRB2020,GharaviPRB2016,AasenPRX2016,KarzigPRB2017}.
We follow the scheme described in \onlinecite{SzechenyiPRB2020}:
conversion is performed by switching on the two tunnel couplings $u_\text{R}$ for an appropriate duration, and reading out the charge of the readout dot afterwards. The tunnel pulse converts the fermion-number parity of the Y junction to the charge of the dot.
The charge readout is assumed to be perfect, and to yield 0 or 1. 

For parity-to-charge conversion, we use sine-squared tunnel pulses:
\begin{equation}
    u_\text{R}(t) =  u_\text{R,max} \sin^2(\pi t/T_\text{R}),\quad\text{if }0\leq t\leq T_\text{R},
\end{equation}
where $u_\text{R,max}$ is the strength of the parity-to-charge conversion, $T_\text{R}=h/(4u_\text{R,max})$ is the duration for ideal conversion. In our simulations, we use $u_\text{R,max}=0.25v$.
Note that this parity-to-charge conversion scheme is not `protected', in the sense that small perturbations, e.g., on-site energy fluctuations, or errors in the tunnel pulse strength or duration, lead to readout errors without any exponential suppression \cite{SzechenyiPRB2020}. 
(For alternative parity-to-charge conversion schemes, and their error mechanisms, see, e.g., Refs.~\cite{AasenPRX2016,VijayPRB2016,KarzigPRB2017,PluggeNJP2017,SteinerPRR2020,MunkPRR2020,KhindanovSciPost2021}.)

\section{Diabatic errors of a perfect Majorana qubit}
\label{sec:clean}

Before describing the effect of quasistatic disorder, we characterize the braiding-based $\pi/2$ gate in a clean, perfectly controlled system. 
We performed numerical simulations of the protocol described above using the time-dependent Bogoliubov-de Gennes (BdG) formalism. 
Details are described in Appendix \ref{app:bdg}.
The BdG formalism allows for efficient simulations: for example, the many-body Fock space of the 8-site setup in Fig.~\ref{fig:setup}a has dimension $2^8 = 256$, growing exponentially with increasing system size $N_\text{c}$, whereas the space of BdG wave functions has dimension $2 \cdot 8 = 16$, growing linearly with increasing system size $N_\text{c}$. 

We solve the time-dependent BdG-Schr\"odinger equation numerically, by discretizing the time axis, approximating the time-dependence of the parameters by step-like dependence, and evaluating the propagator for each time step by exponentiating the instantaneous BdG Hamiltonian matrix. 
Recall that all of our control pulses have sine-squared shape. 
We call each half-period of each sine-squared pulse an \emph{elementary step} in our protocol. 
For each elementary step, having duration $T$, we choose the time discretization step as $\Delta t = \text{min}\{\hbar/v,T/150\}$.

\begin{figure*}
	\centering
	\includegraphics[width=17.8cm]{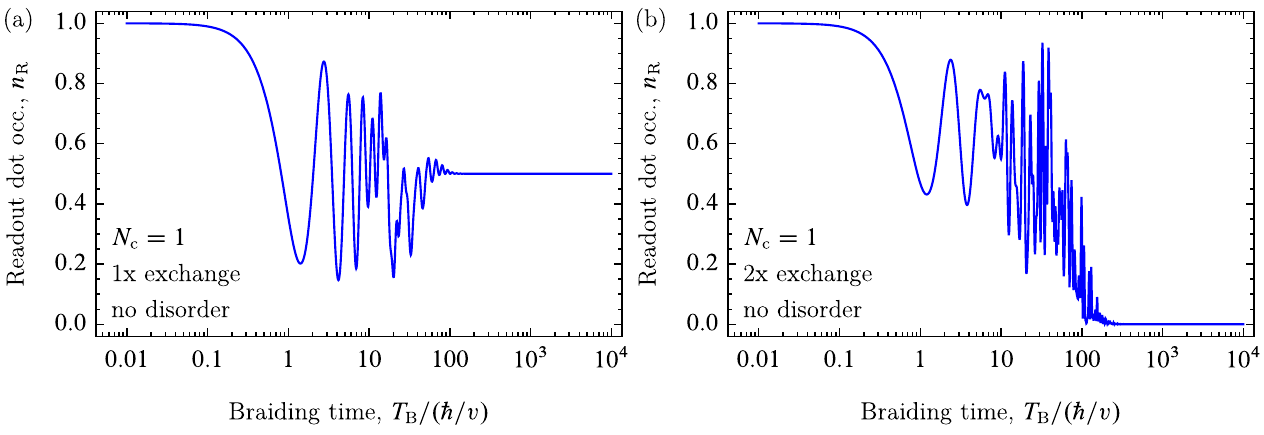}
	\caption{
	In a disorder-free system, a braiding-based $\pi/2$ gate implies a fidelity plateau for long braiding times.
	Readout dot occupation is shown as a function of the braiding time, for the setup with 8 quantum dots ($N_\text{c}=1$), obtained from our numerical simulation of a disorder-free system. 
	(a) Protocol with one exchange of the Majorana zero modes in the Y-junction. (b) Protocol with two exchanges. 
	For short braiding times, braiding does not affect the dynamics resulting in $n_\text{R}\approx1$. For intermediate braiding times, diabatic errors dominate the data. In the adiabatic limit ($T_\text{B} \to \infty$), readout dot occupation shows a fidelity plateau at (a) 1/2 (b) 0, as predicted in Table \ref{tab:idealprotocol}.
	}
	\label{fig:BraidingNodisorder}
	\centering
\end{figure*}

In our numerical experiment, we aim to characterize the braiding-based $\pi/2$ gate via the statistics of the measurement of the readout dot charge. 
This is a realistic constraint: in a real experiment, the experimenter can indeed perform a charge measurement, but has no access to the many-body wave function (let alone any of the BdG wave functions). 
Hence, the target quantity of our simulations is the readout dot occupation, $n_\text{R} = \braket{\Psi_\text{f}|c^\dag_\text{R} c_\text{R}| \Psi_\text{f}}$, where $\Psi_\text{f}$ is the final state, i.e., the many-body state of the system after the parity-to-charge conversion.

The numerical result for this readout dot occupation, for a protocol with a braiding containing $1\times$ exchange, is shown in Fig.~\ref{fig:BraidingNodisorder}a as the function of braiding time $T_\text{B}$. 
This result corresponds to the smallest system size $N_\text{c}=1$, i.e., to the 8-site setup shown in Fig.~\ref{fig:setup}a. 
For short braiding times, $T_\text{B} \lesssim 0.1 \hbar/v$, the exchange protocol is so fast that the wave function is almost unchanged during the braiding because the system has no time to respond to the time-dependence of the Hamiltonian. 
Thus in this limit, the final wave function is close to the one corresponding to no exchange (see `$0\times$ exchange' column in Table \ref{tab:idealprotocol}). 
In the adiabatic limit, however, i.e. for $T_\text{B} \gtrsim 100 \hbar/v$, the readout dot occupation shows a straight `fidelity plateau' at $n_\text{R} = 1/2$, which is consistent with the expectation that braiding induces a $\pi/2$ rotation around the qubit z axis (see `$1\times$ exchange' column in Table \ref{tab:idealprotocol}). 
For intermediate braiding times, the occupation of the readout dot depends strongly on the actual value of the braiding time. Here the MZM exchange is slow enough to induce dynamics in the system, but the diabatic errors are significant.

One might wonder whether an average dot occupation of $n_\text{R} = 1/2$ is a satisfying signature of a precisely functioning quantum gate? As we will see in Sec.~\ref{sec:disorder}, a similar result of $n_\text{R} \approx 1/2$ arises also if the experiment is dominated by strong decoherence that randomizes the Majorana qubit during the exchange. 

Hence we study, as an important control experiment, the case where braiding consists of two consecutive counterclockwise exchanges of the MZMs ($2\times$ exchange). 
Our numerical simulation of this control experiment is shown in Fig.~\ref{fig:BraidingNodisorder}b. Here the braiding time $T_\text{B}$ incorporates both exchanges: $T_\text{B}=12N_\text{c}T_\text{ramp }$. 
The behaviour of the final readout-dot occupation for short and intermediate braiding times is similar to that seen in panel (a). 
However, in the adiabatic limit $T_\text{B} \to \infty$, the readout dot occupation is zero as anticipated, e.g., in the last column of the Table~\ref{tab:idealprotocol}.

\begin{figure}[!h]
	\centering
	\includegraphics[width=\columnwidth]{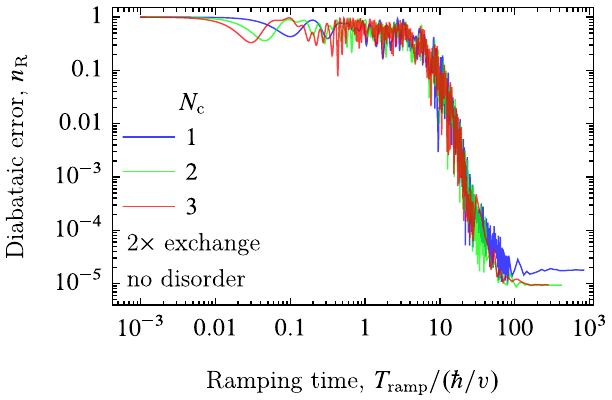}
	\caption{Diabatic errors of the braiding-based $\pi/2$ gate scale with the ramping time.
	Diabatic error is illustrated as a function of the ramping time in the case of the $2\times$ exchange protocol, for three different chain lengths. 
	For this protocol, diabatic error corresponds to the readout dot occupation at the end of the protocol. The three curves overlap, showing that the diabatic error scales with the ramping time (which is inversely proportional to the domain-wall speed), irrespective of the chain length. For longer ramping times $T\gtrsim 100 \hbar /v$, the $n_\text{R}$ curves saturate due to the nonzero error of tunnel-pulse-based gates and readout.
	}
	\label{fig:BraidingRampingTime}
	\centering
\end{figure}

One aspect of the topological protection of a Majorana qubit is that errors induced by the finite overlap of the MZMs can be exponentially suppressed by increasing the system size. 
Even though the simulations in Fig.~\ref{fig:BraidingNodisorder} are not subject to such errors, we highlight an interesting and potentially useful aspect of chain length dependence here.
The length dependence of the final-state readout dot occupation is shown in Fig.~\ref{fig:BraidingRampingTime}. 
Importantly, the horizontal axis shows the time $T_\text{ramp}$ needed to move the topological-trivial domain wall by a single site, 
and not the complete braiding time $T_\text{B} = 12 N_\text{c} T_\text{ramp}$.
Our conclusion is that the results for the three different system sizes $N_\text{c} \in \{1, 2, 3\}$ show very similar power-law-type behavior for intermediate times, which provides the clear prediction that this behavior is universal, and hence can be used to extrapolate for larger system sizes ($N_\text{c} >3$) as well.
This result is analogous to our earlier result for braiding in the SSH model, see Fig. 3b of Ref.~\onlinecite{BorossPRB2019}.

Note also that in Fig.~\ref{fig:BraidingRampingTime}, the apparent error of the gate, i.e., the final occupation of the readout dot, saturates for large ramp times at a plateau of $n_\text{R} \approx 10^{-5}$.
This is a consequence of the non-protected nature of the tunnel-pulse-based $\pi/2$ gates and readout; errors caused by such tunnel-pulse-based operations set the height of this plateau.
(See, e.g., Fig.~2 of \onlinecite{SzechenyiPRB2020} describing readout error in a similar readout scheme.)
This is a feature that we expect to see in future braiding experiments as well.

\section{Charge noise induces Majorana qubit dephasing if braiding is slow}
\label{sec:disorder}

The braiding-based $\pi/2$-gate on the Majorana qubit is an example of a topological quantum gate: the operation on the qubit does not depend on the geometry of the path of control parameters, only on the topology of the path. 
This topological nature implies robustness, i.e., resilience to imperfections in the control path. 
We investigate this robustness here. 
As the imperfection, we focus on $1/f$ charge noise, which is known to be an important ingredient in quantum dots \cite{MishmashPRB2020,CywinskiPRB2008,DialPRL2013,YonedaNatNano2018,ShnirmanPhysScri2002,FreemanAPL2016,HetenyiPRB2019,CywinskiPRB2020}, and hence is expected to be relevant in future Majorana qubit experiments based on quantum dot arrays as well. 

Following earlier studies, we model $1/f$ charge noise in our multi-dot system as \emph{quasistatic disorder} \cite{SzechenyiPRB2020,BorossPRB2022}. 
 We assume that we set our system in the topological fully dimerized limit, but there is also an unwanted, uncontrolled, and spatially uncorrelated random quasistatic contribution to each on-site energy (except to the potential of the readout dot, see below). 
 These on-site energy contributions are  represented as independent zero-mean Gaussian random variables:
\begin{subequations}
\label{eq:gaussiandisorder}
\begin{align}
    \delta\mu_{n,i},\delta\mu_{k} & \sim \mathcal{N}(0,\sigma_\mu^2), \\
    \delta\mu_\text{R} & = 0,
\end{align}
\end{subequations}
where $\sigma_\mu$ is the disorder strength, i.e the standard deviation of the Gaussian distribution, $n=\{1,2,3\}$, $i=\{1,2,...,N_\text{c}\}$ and $k=\{0,1,2,...,2N_\text{c}+1\}$.
Note that static disorder is a much studied ingredient of the Kitaev chain model, see, e.g., Refs.~ \onlinecite{BrouwerPRL2011,HegdePRB2016}. 

In the Ramsey-type protocol we consider in this work, a single data point (readout dot occupation) is obtained by taking multiple runs of the very same experiment and averaging the binary values (0 or 1) of the readout dot charge measured upon the multiple runs. 
The assumption of the quasistatic model is that this averaging procedure is equivalent to averaging over the disorder configurations described by Eq.~\eqref{eq:gaussiandisorder}.

We focus on the effect of disorder on the braiding-based gates. 
Hence, we do not add disorder to the on-site energy of the readout dot, and do not include imperfections in the tunnel pulses $u_\text{P}(t)$ or $u_\text{R}(t)$. 
Such imperfections would in fact cause errors that are not suppressed by increasing system size $N_\text{c}$.
In contrast, the on-site energy disorder (whose effects we describe below) and disorder in the static tunnel amplitudes and pair potentials (which we do not describe explicitly below) cause errors that are suppressed by increasing system size $N_\text{c}$.

\begin{figure*}
	\centering
	\includegraphics[width=17.8cm]{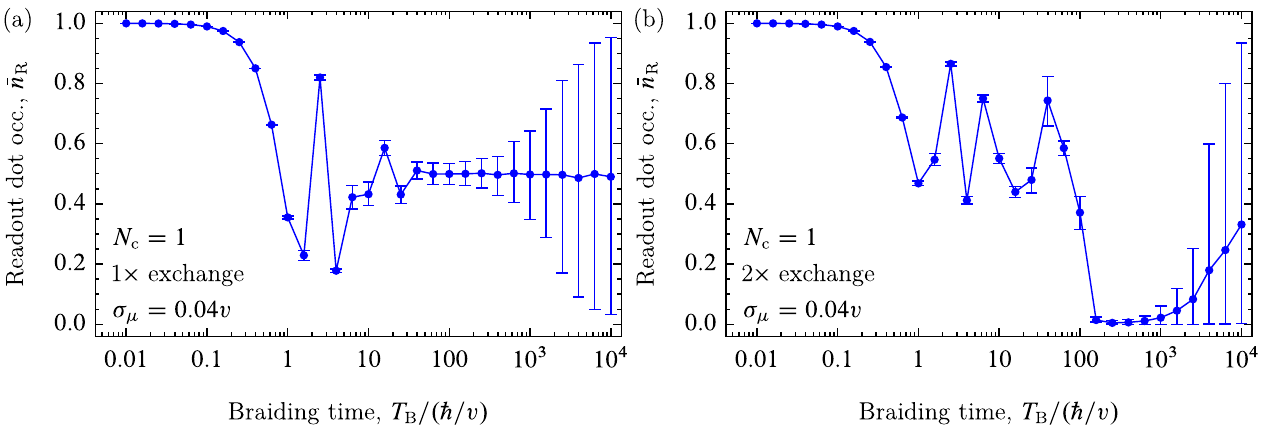}
	\caption{
	Numerical simulation of the braiding experiment in the presence of on-site disorder. 
    Disorder-averaged final-state occupation $\bar{n}_\text{R}$ of the readout dot is shown as a function of the braiding time, for the smallest system size $N_c = 1$ with disorder strength $\sigma_\mu = 0.04$. 
    Each curve is a result of $N_\text{r} = 1000$ realizations, dots show the mean of the occupations, while error bars depict the $10\%-90\%$ quantiles. (a) Single exchange. 
    Disorder-averaged readout dot occupation for long braiding time is similar to that without disorder (see Fig.~\ref{fig:BraidingNodisorder}a), but the error bars arise due to disorder. 
    (b) Double exchange. The disorder changes the mean of the readout dot occupation in the adiabatic limit (cf.~Fig.~\ref{fig:BraidingNodisorder}b), and the error bars also show the influence of disorder. The plateau between $100$ and $1000$ can serve as the signature of  successful braiding in a real experiment.
	}
	\label{fig:BraidingDisorderErrorbar}
	\centering
\end{figure*}

Figure \ref{fig:BraidingDisorderErrorbar}a shows the disorder-averaged result of the Ramsey-type protocol with $1\times$ exchange, averaging $N_\text{r} = 1000$ different disorder realizations for the disorder strength $\sigma_\mu=0.15v$. 
We show results of simulations where the initial state is formed by the product of the even ground state of the Y junction, the even ground state of the wire, and the empty readout dot (see App.~\ref{app:bdg} for details). 
The  quantity we plot in Fig.~\ref{fig:BraidingDisorderErrorbar}a is the disorder-averaged final readout dot occupation 
\begin{equation}
    \bar{n}_\text{R}
    = \frac{1}{N_\text{r}}
    \sum_{j=1}^{N_\text{r}} 
    n_\text{R}^{(j)},
\end{equation}
where $n_\text{R}^{(j)}$ is the expectation value of the final readout dot occupation for the $j$th disorder realization.
Furthermore, the error bars show the $10\%-90\%$ quantiles of the 1000 different realizations. 

Figure \ref{fig:BraidingDisorderErrorbar}a shows that for short braiding times, $T_\text{B} \lesssim 50 \hbar /v$, diabatic errors corrupt the gate. 
For intermediate times, $50 \hbar /v \lesssim T_\text{B} \lesssim 100 \hbar /v$, the readout dot occupation is close to the clean result $n_\text{R}=1/2$, with small fluctuations as shown by the error bars. 
For long times, $T_\text{B} \gtrsim 100 \hbar /v$, the readout dot occupation is still close to the clean result $n_\text{R} = 1/2$, but the fluctuations grow significantly.

The fluctuations seen for $T_\text{B} \gtrsim 50 \hbar/v$ in  Fig.~\ref{fig:BraidingDisorderErrorbar}a are interpreted as dephasing occuring during the braiding-based quantum operation. 
The on-site energy disorder detunes the Kitaev chains from the fully dimerized limit, causing hybridization of MZMs and corresponding energy splittings, both in the Y junction and in the straight Kitaev wire. As a consequence of the splittings, the dynamical phases acquired by the Majorana qubit basis states become different, which in turn causes significant deviations from the ideal (disorder-free and adiabatic) scenario, which relies only on the geometric phase difference $\pi/2$ between the even and odd ground states of the Y junction.

Although the increasing error bars with increasing braiding time, seen 
in  Fig.~\ref{fig:BraidingDisorderErrorbar}a
for $T_\text{B} \gtrsim 50 \hbar/v$, provide a clear numerical signature of dephasing, these error bars cannot be revealed by an experiment that follows our protocol. 
The reason is as follows. 
The randomness of disorder implies that the final dot occupation $n_\text{R}$ itself is a random variable.
However, in a single run of the experiment, say, the $j$th run, the charge readout result is either 0 or 1, with the probability of 1 determined by $n^{(j)}_\text{R}$.
The data point for a given braiding time is the average of these binary results for a large number of runs with a fixed $T_\text{B}$.
This averaging procedure yields the disorder-averaged $\bar{n}_\text{R}$, and the result carries no information about the disorder-induced \emph{fluctuations} of $n_\text{R}$.

Nevertheless, dephasing during the braiding-based gate can be characterised by our protocol. 
This is done by doing the exchange twice.
In that setting, the idealized readout dot occupation is $n_\text{R} = 0$, and dephasing causes deviations of the disorder-averaged $\bar{n}_\text{R}$ from that value.  
Fig.~\ref{fig:BraidingDisorderErrorbar}b shows the disorder-averaged result of the $2\times$ exchange protocol, along with error bars as discussed above. 
For short braiding times $T_\text{B} \lesssim 200 \hbar /v$, there is no significant difference from the clean case, and the result is dominated by diabatic error. 
For intermediate braiding times, $ 200 \hbar /v \lesssim T_\text{B} \lesssim 1000 \hbar/v$, a `fidelity plateau' is seen, which is the signature that each MZM exchange realizes a $\pi/2$ gate.
We anticipate that such fidelity plateaus will serve as important signatures of topologically protected quantum gates in future experiments.
For longer braiding times, $T_\text{B} \gtrsim 1000 \hbar /v$, the result is dominated by disorder-induced dephasing that happens during the MZM exchanges.  

\begin{figure*}
	\centering
	\includegraphics[width=17.8cm]{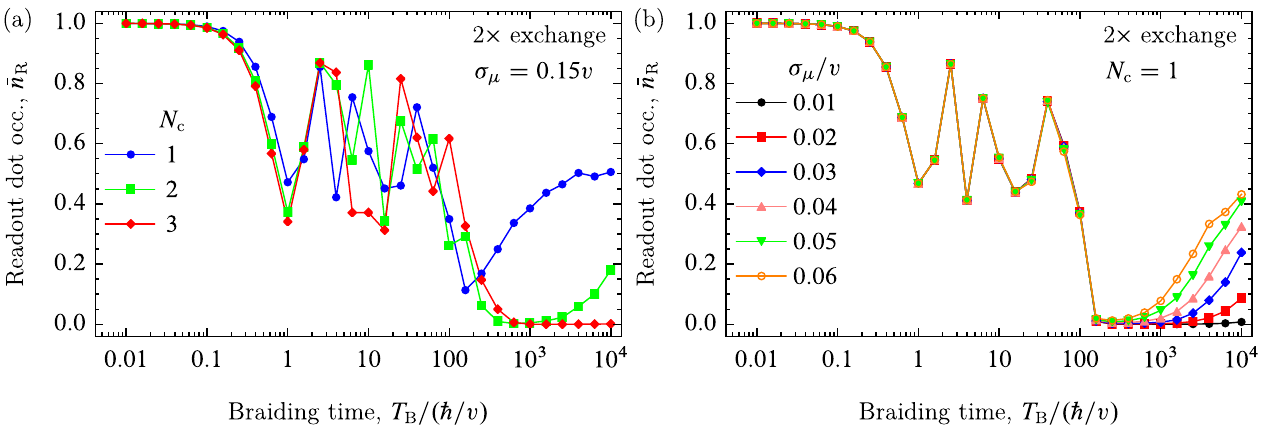}
	\caption{
	Demonstration of topological protection. 
    Disorder-averaged occupation of the readout dot as a function of the braiding time is shown in the double-exchange protocol (a) for a given disorder strength $\sigma_\mu = 0.15 v$ and for different system sizes; (b) for the smallest system size $N_\text{c}=1$ and for different disorder strengths. 
    Each curve is obtained by averaging for $N_\text{r} = 1000$ realizations. 
    For (a) increasing system size, and (b) for decreasing disorder strength, a plateau with $\bar{n}_\text{R} \approx 0$ appears for long braiding times, and the plateau length increases.}
	\label{fig:BraidingDisorderMean}
	\centering
\end{figure*}

To illustrate the topological protection of the braiding-based quantum gate, we show  how the fidelity plateau length varies if we vary the system size and the strength of the disorder. 
In Fig.~\ref{fig:BraidingDisorderMean}a, the disorder-averaged readout dot occupation $\bar{n}_\text{R}$ is shown for the three smallest system sizes $N_\text{c}=1,2,3$ using the $2\times$ exchange protocol. 
Disorder strength is set to $\sigma_\mu = 0.15 v$, which is strong enough to reduce the plateau to a dip for $N_\text{c}=1$ (blue). 
By increasing the chain length to $N_\text{c}=2$ (green), a fidelity plateau is developed, which flattens further for an even larger system size $N_\text{c} = 3$ (red). 
This improvement of the gate quality by increasing the system size will probably be used as a hallmark of topologically protected quantum gates in future braiding experiments.
In Fig.~\ref{fig:BraidingDisorderMean}b, we show the disorder-averaged readout dot occupation curves for different disorder strengths in the case of the smallest system size $N_\text{c}=1$. The main observation here is that the fidelity plateau gets longer and more flat as disorder strength is decreased.  

\section{Discussion}
\label{sec:discussion}

\subsection{Experimental time scales}
\label{sec:exptimescales}

Here, we convert our numerical results discussed above to quantitative estimates of time scales for future experiments. 
Based on the recent experiment of Ref.~\onlinecite{Dvir}, we assume an induced gap of $30\, \mu$eV in our fully dimerized Kitaev-chain model, equivalent to setting $v = \Delta = 15 \, \mu$eV.
The simplest quantum-dot-based braiding setup corresponds to our $N_\text{c}=1$ case, requiring 8 quantum dots. 

Assuming that our protocol will be realized experimentally in such a setup, we ask the question: for which experimental result are we convinced that the experiment demonstrated a braiding-based topological quantum gate? 
That is, when do we say that a fidelity plateau is `long and flat' enough to be a convincing evidence of such a gate? 
To make this question a bit more specific: what is the level of disorder that enables the experimenter to observe a fidelity plateau below the error level of one percent ($\bar{n}_\text{R} \leq 0.01$) such that the fidelity plateau spreads over at least one order of magnitude along the braiding-time axis?
Using the estimate $v = \Delta = 15 \, \mu$eV, a quantitative answer can be read off Fig.~\ref{fig:BraidingDisorderMean}b:
for $\sigma_\mu = 0.02 v = 0.3 \, \mu$eV (red squares), the fidelity plateau spreads in the range
$ 250 \hbar / v \lesssim 
T_\text{B} \lesssim 
2500\hbar/v$,  
which is equivalent to 
$5.5\text{ ns}
\lesssim
T_\text{B}
\lesssim
55\text{ ns}$.

We note that typical quantum dot devices show on-site energy fluctuations ($\sigma_\mu$) of the order of a few microelectronvolts, see, e.g., the  experimental data listed in Table II of Ref.~\onlinecite{KnappPRB2018}. However, in recent state-of-the-art experiments with a double quantum dot, $\sigma_\mu \sim 0.1 \, \mu$eV has been achieved\cite{ScarlinoArxiv2021}.
Even though it is a highly nontrivial technological challenge to combine the high-quality dot structures with superconductors, these numbers give hope that topological quantum gates can be experimentally demonstrated using small quantum dot arrays.

\subsection{Beyond the minimal model}

The model we have used in this work is a minimal one, allowing us to focus on a few key physical ingredients, and to highlight a few mechanisms likely to affect the proposed experiment. 
Here, we list further ingredients worthwhile to incorporate in future studies.
\begin{enumerate}
    \item \emph{Modelling braiding dynamics with a microscopic dot-array Hamiltonian.}
    We have assumed that the Kitaev chain model provides a faithful description of a quantum dot array proximitised by superconductors. 
    Much physical insight behind that assumption has been provided by Refs.~\onlinecite{Leijnse_2012,SauNatComm2012,Fulga_2013,Tsintzis}. 
    However, it will be an interesting and relevant extension of our work to study the braiding-induced dynamics in a more realistic dot array model, where, e.g., Zeeman sublevels of the dots, excited orbitals, Coulomb repulsion, spin-orbit interaction, etc, are explicitly incorporated.
    \item \emph{Role of fluctuations of $v$ and $\Delta$.}
    We have assumed that the static tunnel amplitudes ($v$) and pair potentials ($\Delta$) are perfectly controlled. In an experiment, this is not the case. For small quantum dot arrays, the effect of their fluctuations can be relevant. For example, the MZM energy splitting of a two-site Kitaev chain is linear in the parameter $|v| - |\Delta|$ (see Ref.~\onlinecite{Leijnse_2012}), hence the fluctuations of $v$ and $\Delta$ can cause significant dephasing. 
    Note that this dephasing mechanism is suppressed exponentially by increasing the system size, similar to the case of on-site energy fluctuations we studied above. 
    \item \emph{Errors in the tunnel pulses.} A further difference between our minimal model and experiments is that in the latter, the tunnel pulses are imperfect; e.g., pulse duration, pulse amplitude, pulse shape deviate from ideal.
    Such perturbations, if strong enough, could significantly reduce the quality (flatness and length) of the fidelity plateaus shown in Figs.~\ref{fig:BraidingDisorderErrorbar}b and \ref{fig:BraidingDisorderMean}.
    \item \emph{On-site energy fluctuations of the readout dot.}
    In this study, we have disregarded on-site energy fluctuations of the readout dot, to focus on the features of MZM braiding. 
    Such fluctuations lead to readout error\cite{SzechenyiPRB2020}, which, similarly to the effect of tunnel pulse errors, reduces the quality of the fidelity plateaus.
    Based on our earlier results\cite{SzechenyiPRB2020}, we estimate that extending the on-site energy fluctuations to the readout dot would not change the quantitative time-scale analysis of Sec.~\ref{sec:exptimescales}.
 
    \item \emph{Beyond the quasistatic approximation.}
    In this work, we use the quasistatic approximation\cite{TosiNatComm2017,BoterPRB2020} to describe the effect of charge noise.
    In real devices, charge noise often exhibits a $1/f$-type spectrum\cite{KnappPRB2018,MishmashPRB2020,DialPRL2013,YonedaNatNano2018,FreemanAPL2016,HetenyiPRB2019,CywinskiPRB2020,ShnirmanPhysScri2002,MakhlinNewDirections2003,HuangPRA2019,KhindanovSciPost2021}.
    Incorporating the $1/f$ frequency dependence of noise in our model, following, e.g., Refs.~\onlinecite{MakhlinNewDirections2003,AasenPRX2016,MishmashPRB2020,KhindanovSciPost2021},
    could be an important addition to the present work. 
    
    \item \emph{Qubit initialization.}
    Above, we assumed that the Majorana qubit can be initialized to a particular state of the four-fold degenerate ground-state manifold.
    One way to achieve this experimentally is via thermalization. 
    The experimenter can tune both the Y junction and the straight wire away from the fully dimerized limit, to open up an energy splitting between the even and odd ground states of both subsystems, and to make the ground state unique (e.g., $\ket{\text{e},\text{e}}$). 
    Then, thermalization will relax the system to this unique ground state, completing the initialization step.

    \item \emph{Thermalization-induced decoherence during braiding.} In our simulations above, we neglect effects and errors induced by the finite-temperature bath that is unavoidably present in a real experiment. 
    This effect is discussed in detail in Ref.~\onlinecite{Breckwoldt_2022}, using a Kitaev-chain model similar to ours. 
    For a simple numerical estimate based on the parameters of Sec.~\ref{sec:exptimescales}, we assume that the thermalizing bath cannot change the fermion number parity of the multi-dot setup during the braiding phase.  
    For that case, we estimate the error due to thermalization as the probability of having excitation energy $2v$ in thermal equilibrium.
    This error has no significant effect on the fidelity plateau defined by $n_\text{R} \leq 0.01$, if $e^{-2v / k_\text{B} T} \leq 0.01$, which is converted to the condition $T \leq 76\, \text{mK}$ for $v = 15\, \mu$eV as assumed in Sec.~\ref{sec:exptimescales}.
    Such temperatures are achieved in dilution refrigerators. 
\end{enumerate}

\section{Conclusions}
\label{sec:conclusions}

In conclusion, we have proposed a setup and a protocol for experimental demonstration of a braiding-based $\pi/2$ gate on a Majorana qubit. The protocol is based on a proximitized quantum-dot array, and it is composed of auxiliary (non-topological) quantum gates, a braiding-based topological quantum gate, and qubit readout via parity-to-charge conversion and charge measurement. 
We focused on the effect of charge noise, which we incorporate in our simulations as quasistatic on-site disorder. 
Our results confirm that the braiding-based gate suffers from diabatic errors for short braiding times, and demonstrate noise-induced dephasing for long braiding times. 
For intermediate times, a fidelity plateau can develop, which is made flatter and longer if disorder strength is decreased or if system size is increased. 
Our numerical results provide quantitative predictions for the quality of future braiding experiments, that can hopefully be built by combining today's state-of-the-art quantum dot arrays and proximity-induced superconductivity.

\acknowledgments

We thank J.~Asb\'oth and G.~Sz\'echenyi for useful discussions. 
This research was supported by the Ministry of Culture and Innovation and the National Research, Development and Innovation Office (NKFIH) within the Quantum Information National Laboratory of Hungary (Grant No. 2022-2.1.1-NL-2022-00004), by the NKFIH via the OTKA Grant No. 132146, and
by the European Union within the Horizon Europe research and innovation programme via the project `IGNITE'.

\appendix

\section{Bogoliubov-de Gennes (BdG) formalism for dynamics}
\label{app:bdg}

Our numerical results in the main text are obtained by applying the BdG formalism. Here we outline how to 
calculate the time dependence of the readout dot occupation via
\begin{equation}
\label{eq:chargeevolution}
    n_\text{R}(t) = \braket{\Psi(t)|c_\text{R}^\dag c_\text{R}^{\vphantom{\dag}}|\Psi(t)},
\end{equation}
where $\ket{\Psi(t)}$ is the time-evolved many-body state developing from the initial state $\Psi_\text{i}$ of the system.

The time-dependent Fock-space Hamiltonian in Eq.~\eqref{eq:fullFockHamiltonian} describes the time evolution of the system. To construct the BdG Hamiltonian, we define the vector of the local fermionic operators
\begin{equation}
    \label{eq:ctilde}
    \tilde{\vec{c}} = \begin{pmatrix}
                            \vec{c}_\text{Y} \\
                            \vec{c}_\text{Y}^\dag \\
                            \vec{c}_\text{W} \\
                            \vec{c}_\text{W}^\dag \\
                            c_\text{R} \\
                            c_\text{R}^\dag \\
                      \end{pmatrix},
\end{equation}
where $\vec{c}_\text{Y}$ and $\vec{c}_\text{W}$ denote the vector of annihilation operators of the Y-junction and the straight wire, respectively.
Using the vector $\tilde{\vec{c}}$, we can rewrite the Hamiltonian as
\begin{equation}
    H(t) = \frac{1}{2}\tilde{\vec{c}}^\dag\mathcal{H}(t)\tilde{\vec{c}} + \frac{1}{2}\sum_{i\in\text{L}}{\mu_i},
\end{equation}
where $\mathcal{H}(t)$ is the BdG Hamiltonian, and $L$ is the set of all site labels.

At $t=0$, the system is decoupled to three components (Y-junction, straight wire, readout dot) by setting $u_\text{R}(0)=u_\text{P}(0)=0$, and hence the BdG Hamiltonian has the form 
\begin{equation}
    \mathcal{H}(0) =
    \begin{pmatrix}
         \mathcal{H}_\text{Y}(0) & 0 & 0\\
         0 & \mathcal{H}_\text{W}(0) & 0\\
         0 & 0 & \mathcal{H}_\text{R}(0)
    \end{pmatrix},
\end{equation}
where $\mathcal{H}_\text{Y}(0)$, $\mathcal{H}_\text{W}(0)$ and $\mathcal{H}_\text{Y}(0)$ are the BdG Hamiltonian of the Y-junction, the straight wire and the readout dot, at $t=0$, respectively.

As usual in the BdG formalism, we use a time-dependent BdG Hamiltonian that is particle-hole symmetric for all times $t$:
\begin{equation}
    \mathcal{P}\mathcal{H}(t)\mathcal{P}^{-1} = - \mathcal{H}(t).
\end{equation}
Here, the particle-hole transformation reads
\begin{equation}
    \mathcal{P} = \begin{pmatrix}
         \sigma_x \otimes \mathbb{1}_{N_\text{Y}} & 0 & 0\\
         0 & \sigma_x \otimes \mathbb{1}_{N_\text{W}} & 0\\
         0 & 0 & \sigma_x
    \end{pmatrix}K,
\end{equation}
where $\sigma_x$ is the first Pauli matrix acting on the Nambu (particle-hole) degree of freedom,  $\mathbb{1}_n$ is the $n \times n$ identity matrix, $N_\text{Y}$ ($N_\text{W}$) is the number of the sites in the Y-junction (straight wire), and $K$ is complex conjugation. We note that each of the three subsystems alone has particle-hole symmetry.

Our goal is to describe the expectation value of the readout dot occupation 
$c^\dag_\text{R} c^{\vphantom{\dag}}_\text{R}$
at the final moment of the protocol. 
To this end, we first solve (numerically) the eigenvalue problem of the initial BdG Hamiltonians, $\mathcal{H}_\text{Y}(0)$, $\mathcal{H}_\text{W}(0)$ and $\mathcal{H}_\text{Y}(0)$, we find the eigenvalues of the subsystems $\lambda_{\text{Y},i}$, $\lambda_{\text{W},i}$ and $\lambda_{\text{R},i}$, as well as the corresponding eigenvectors $\phi_{\text{Y},i}$, $\phi_{\text{Y},i}$ and $\phi_{\text{Y},i}$, where $i=1,2,...,2N_k$, $N_k$ is the number of sites in the given subsystem and $k\in\{\text{Y},\text{W},\text{R}\}$ denotes the subsystem. 
We order the eigenvalues such that $\lambda_{k,i} \geq 0$ and $\lambda_{k,i+N} = -\lambda_{k,i}$ for $1\leq i \leq N \equiv N_\text{Y}+N_\text{W}+N_\text{R} $. 
Furthermore we choose the eigenvectors such that $\phi_{k,i+N} = \mathcal{P} \phi_{k,i}$.

With the eigenvectors $\phi_{k,i}$ at hand, we can express the unitary matrix $U'$ that diagonalizes $\mathcal{H}(0)$:
\begin{equation}
    U' =
    \begin{pmatrix}
         U_\text{Y} & 0 & 0\\
         0 & U_\text{W} & 0\\
         0 & 0 & U_\text{R}
    \end{pmatrix},
\end{equation}
where $U_\text{Y}$, $U_\text{W}$ and $U_\text{R}$ are the diagonalizers of the Y-junction, the straight wire and the readout dot, respectively. 
The diagonalizer of subsystem $k \in \{\text{Y}, \text{W}, \text{R}\}$ can be written as
\begin{equation}
    U_k =
    \begin{pmatrix}
         \phi_{k,1}^\dag\\
         ...\\
        \phi_{k,2N_k}^\dag
    \end{pmatrix}.
\end{equation}
The vector of quasiparticle operators can be expressed as
\begin{equation}
\label{eq:dprime}
    \tilde{\vec{d}}' = U'  \tilde{\vec{c}} =
                        \begin{pmatrix}
                            \vec{d}_\text{Y} \\
                            \vec{d}_\text{Y}^\dag \\
                            \vec{d}_\text{W} \\
                            \vec{d}_\text{W}^\dag \\
                            d_\text{R} \\
                            d_\text{R}^\dag
                        \end{pmatrix},
\end{equation}
where $\vec{d}_\text{Y}$, $\vec{d}_\text{W}$ and $d_\text{R}$ are the quasiparticle operators corresponding to the Y-junction, the straight wire and the readout dot, respectively.
The fact that the on-site energy of the readout dot is zero leaves $d_\text{R}$ ambiguous; for concreteness, we define $d_\text{R} = c_\text{R}$.

To highlight the special role of the quasi-zero-energy excitations, we reorder the vector of quasiparticle operations as
\begin{equation}
    \label{eq:dUc}
    \tilde{\vec{d}} = \underbrace{\Pi U'}_{U}  \tilde{\vec{c}} =
                        \begin{pmatrix}
                            d_1 \\
                            ... \\
                            d_N \\
                            d_1^\dag \\
                            ... \\
                            d_N^\dag
                        \end{pmatrix},
\end{equation}
where $d_1$ ($d_2$) corresponds to the quasi-zero-energy excitation of the Y junction (straight wire), $d_i\,(i=3,...N)$ are ordered such a way that the corresponding excitation energies are in ascending order, furthermore $\Pi$ is the permutation matrix corresponding to the reordering.

As stated above, our goal is to compute the time dependence of the readout dot occupation via Eq.~\ref{eq:chargeevolution}, which we rephrase as 
\begin{equation}
    \label{eq:readoutdotoccupation}
    n_\text{R}(t) = \braket{\Psi_\text{i}|[\tilde{\vec{c}}(t)]_{2N} [\tilde{\vec{c}}(t)]_{2N-1}|\Psi_\text{i}},
\end{equation}
where $[...]_j$ denotes the $j$th component of the vector, $\ket{\Psi_\text{i}}$ is a given initial state of the system, and  the elements of the vector $\tilde{\vec{c}}(t)$ are the elements of the vector $\tilde{\vec{c}}$, transformed to the Heisenberg picture:
\begin{equation}
    \label{eq:cheisenberg}
    \tilde{\vec{c}}(t) = \underbrace{\mathcal{T} e^{-\frac{i}{\hbar}\int_{0}^{t}{\mathcal{H}(\tau)d\tau}}}_{\mathcal{U}(t)} \tilde{\vec{c}},
\end{equation}
where $\mathcal{T}$ is the time ordering operator and $\mathcal{U}(t)$ is the BdG propagator.
Inserting Eq.~\eqref{eq:cheisenberg} into Eq.~\eqref{eq:readoutdotoccupation} and using Eq.~\eqref{eq:dUc}, we obtain
\begin{align}
    n_\text{R}(t) &= \braket{\Psi_\text{i}|[\mathcal{U}U^\dag  \tilde{\vec{d}}]_{2N}[\mathcal{U}U^\dag  \tilde{\vec{d}}]_{N}|\Psi_\text{i}} \nonumber \\
    &= \sum_{n,m=1}^{2N}{S_{nm}(t)\braket{\Psi_\text{i}|\tilde{d}_n \tilde{d}_m|\Psi_\text{i}}},
\end{align}
where the second line is an implicit definition of 
\begin{equation}
    S_{nm}(t) = [\mathcal{U}U^\dag]_{2N,n}[\mathcal{U}U^\dag]_{N,m}.
\end{equation}

In the main text, we focus on the case when $\ket{\Psi_\text{i}}=\ket{\text{e},\text{e},0}$, i.e., both the Y-junction and the straight wire is in its even ground state, and the readout dot is empty. 
The even-parity state could be the actual ground state (denoted by `G') or the first excited state (within the quasidegenerate ground state subspace, denoted by `E') depending on the actual on-site energy disorder realization. The two state are related as
\begin{subequations}
\begin{align}
    \ket{\text{E},\text{G},0} &= d_1^\dag \ket{\text{G},\text{G},0}, \\
    \ket{\text{G},\text{E},0} &= d_2^\dag \ket{\text{G},\text{G},0}.
\end{align}
\end{subequations}

The readout dot occupation can be calculated for all four initial energy eigenstates
$\ket{\text{G},\text{G},0}$,
$\ket{\text{E},\text{G},0}$,
$\ket{\text{G},\text{E},0}$, and
$\ket{\text{E},\text{E},0}$.
Evaluation of the matrix element of $\braket{\Psi_\text{i}|\tilde{d}_n \tilde{d}_m|\Psi_\text{i}}$ in these four cases yields
\begin{subequations}
\begin{align}
    n_\text{R}(t)\Bigr|_{\ket{\Psi_\text{i}}=\ket{\text{G},\text{G},0}}&=\sum_{\mu=1}^{N}{S_{\mu,N+\mu}(t)}, \\
    n_\text{R}(t)\Bigr|_{\ket{\Psi_\text{i}}=\ket{\text{E},\text{G},0}}&=S_{N+1,1}(t)+\sum_{\mu=2}^{N}{S_{\mu,N+\mu}(t)}, \\
    n_\text{R}(t)\Bigr|_{\ket{\Psi_\text{i}}=\ket{\text{G},\text{E},0}}&=S_{N+2,2}(t)+\hspace{-2em}\sum_{\mu\in\{1,3,4,...,N\}}\hspace{-2em}{S_{\mu,N+\mu}(t)}, \\
    n_\text{R}(t)\Bigr|_{\ket{\Psi_\text{i}}=\ket{\text{E},\text{E},0}}&=S_{N+1,1}(t)+S_{N+2,2}(t)\nonumber\\&+\sum_{\mu=3}^{N}{S_{\mu,N+\mu}(t)}.
\end{align}
\end{subequations}
To determine which case corresponds to the initial state $\ket{\text{e},\text{e},0}$, we use the relation that
\begin{equation}
    \label{eq:evenodd}
    \det(U_{\text{Y}}) =
                            \begin{cases}
                                1,  & \text{if the actual ground state is even},\\
                                -1, & \text{if the actual ground state is odd},
                            \end{cases}
\end{equation}
and the corresponding relation for $U_{\text{W}}$.

Applying Eq.~\eqref{eq:evenodd} for the Y-junction and for the straight wire, the state $\ket{\text{e},\text{e},0}$ can be identified as
\begin{equation}
    \label{eq:actualgroundstate}
    \ket{\text{e},\text{e},0} =
    \begin{cases}
        \ket{\text{G},\text{G},0}, & \text{if }\det(U_{\text{Y}})=\det(U_{\text{W}})=1,\\
        \ket{\text{E},\text{G},0}, & \text{if }-\det(U_{\text{Y}})=\det(U_{\text{W}})=1,\\
        \ket{\text{G},\text{E},0}, & \text{if }\det(U_{\text{Y}})=-\det(U_{\text{W}})=1,\\
        \ket{\text{E},\text{E},0}, & \text{if }\det(U_{\text{Y}})=\det(U_{\text{W}})=-1.
    \end{cases}
\end{equation}

We note that if the on-site energy of the readout dot is zero, or if the on-site disorder is absent in the Kitaev chains, then there is at least one excitation with exactly zero energy. 
This leads to the degeneracy of the zero eigenvalue of $\mathcal{H}_\text{Y}(0)$, $\mathcal{H}_\text{W}(0)$ and $\mathcal{H}_\text{R}(0)$. 
In turn, this leads to an ambiguity in constructing the matrices $U_\text{Y}$, $U_\text{W}$ and $U_\text{R}$. 
We have already fixed the ambiguity of $U_\text{R}$ after Eq.~\eqref{eq:dprime}.
Regarding the Y-junction and straight wire, hosting Majoranas, care must be taken to use \emph{fermionic} zero modes when assembling $U_\text{Y}$ and $U_\text{W}$. Then,  Eq.~\eqref{eq:actualgroundstate} can be used to identify $\ket{\text{e}, \text{e}, 0}$. 

\bibliography{main-v2}

\begin{thebibliography}{81}%
\makeatletter
\providecommand \@ifxundefined [1]{%
 \@ifx{#1\undefined}
}%
\providecommand \@ifnum [1]{%
 \ifnum #1\expandafter \@firstoftwo
 \else \expandafter \@secondoftwo
 \fi
}%
\providecommand \@ifx [1]{%
 \ifx #1\expandafter \@firstoftwo
 \else \expandafter \@secondoftwo
 \fi
}%
\providecommand \natexlab [1]{#1}%
\providecommand \enquote  [1]{``#1''}%
\providecommand \bibnamefont  [1]{#1}%
\providecommand \bibfnamefont [1]{#1}%
\providecommand \citenamefont [1]{#1}%
\providecommand \href@noop [0]{\@secondoftwo}%
\providecommand \href [0]{\begingroup \@sanitize@url \@href}%
\providecommand \@href[1]{\@@startlink{#1}\@@href}%
\providecommand \@@href[1]{\endgroup#1\@@endlink}%
\providecommand \@sanitize@url [0]{\catcode `\\12\catcode `\$12\catcode
  `\&12\catcode `\#12\catcode `\^12\catcode `\_12\catcode `\%12\relax}%
\providecommand \@@startlink[1]{}%
\providecommand \@@endlink[0]{}%
\providecommand \url  [0]{\begingroup\@sanitize@url \@url }%
\providecommand \@url [1]{\endgroup\@href {#1}{\urlprefix }}%
\providecommand \urlprefix  [0]{URL }%
\providecommand \Eprint [0]{\href }%
\providecommand \doibase [0]{http://dx.doi.org/}%
\providecommand \selectlanguage [0]{\@gobble}%
\providecommand \bibinfo  [0]{\@secondoftwo}%
\providecommand \bibfield  [0]{\@secondoftwo}%
\providecommand \translation [1]{[#1]}%
\providecommand \BibitemOpen [0]{}%
\providecommand \bibitemStop [0]{}%
\providecommand \bibitemNoStop [0]{.\EOS\space}%
\providecommand \EOS [0]{\spacefactor3000\relax}%
\providecommand \BibitemShut  [1]{\csname bibitem#1\endcsname}%
\let\auto@bib@innerbib\@empty
%</preamble>
\bibitem [{\citenamefont {Ivanov}(2001)}]{IvanovPRL2001}%
  \BibitemOpen
  \bibfield  {author} {\bibinfo {author} {\bibfnamefont {D.~A.}\ \bibnamefont
  {Ivanov}},\ }\bibfield  {title} {\enquote {\bibinfo {title} {Non-{Abelian}
  statistics of half-quantum vortices in $\mathit{p}$-wave superconductors},}\
  }\href {\doibase 10.1103/PhysRevLett.86.268} {\bibfield  {journal} {\bibinfo
  {journal} {Phys. Rev. Lett.}\ }\textbf {\bibinfo {volume} {86}},\ \bibinfo
  {pages} {268--271} (\bibinfo {year} {2001})}\BibitemShut {NoStop}%
\bibitem [{\citenamefont {Kitaev}(2001)}]{KitaevPU2001}%
  \BibitemOpen
  \bibfield  {author} {\bibinfo {author} {\bibfnamefont {A~Yu}\ \bibnamefont
  {Kitaev}},\ }\bibfield  {title} {\enquote {\bibinfo {title} {Unpaired
  {Majorana} fermions in quantum wires},}\ }\href {\doibase
  10.1070/1063-7869/44/10s/s29} {\bibfield  {journal} {\bibinfo  {journal}
  {Physics-Uspekhi}\ }\textbf {\bibinfo {volume} {44}},\ \bibinfo {pages}
  {131--136} (\bibinfo {year} {2001})}\BibitemShut {NoStop}%
\bibitem [{\citenamefont {Alicea}\ \emph {et~al.}(2011)\citenamefont {Alicea},
  \citenamefont {Oreg}, \citenamefont {Refael}, \citenamefont {von Oppen},\
  and\ \citenamefont {Fisher}}]{AliceaNatPhys2011}%
  \BibitemOpen
  \bibfield  {author} {\bibinfo {author} {\bibfnamefont {Jason}\ \bibnamefont
  {Alicea}}, \bibinfo {author} {\bibfnamefont {Yuval}\ \bibnamefont {Oreg}},
  \bibinfo {author} {\bibfnamefont {Gil}\ \bibnamefont {Refael}}, \bibinfo
  {author} {\bibfnamefont {Felix}\ \bibnamefont {von Oppen}}, \ and\ \bibinfo
  {author} {\bibfnamefont {Matthew P.~A.}\ \bibnamefont {Fisher}},\ }\bibfield
  {title} {\enquote {\bibinfo {title} {Non-{Abelian} statistics and topological
  quantum information processing in {1D} wire networks},}\ }\href {\doibase
  10.1038/nphys1915} {\bibfield  {journal} {\bibinfo  {journal} {Nature
  Physics}\ }\textbf {\bibinfo {volume} {7}},\ \bibinfo {pages} {412--417}
  (\bibinfo {year} {2011})}\BibitemShut {NoStop}%
\bibitem [{\citenamefont {Mourik}\ \emph {et~al.}(2012)\citenamefont {Mourik},
  \citenamefont {Zuo}, \citenamefont {Frolov}, \citenamefont {Plissard},
  \citenamefont {Bakkers},\ and\ \citenamefont
  {Kouwenhoven}}]{MourikScience2012}%
  \BibitemOpen
  \bibfield  {author} {\bibinfo {author} {\bibfnamefont {V.}~\bibnamefont
  {Mourik}}, \bibinfo {author} {\bibfnamefont {K.}~\bibnamefont {Zuo}},
  \bibinfo {author} {\bibfnamefont {S.~M.}\ \bibnamefont {Frolov}}, \bibinfo
  {author} {\bibfnamefont {S.~R.}\ \bibnamefont {Plissard}}, \bibinfo {author}
  {\bibfnamefont {E.~P. A.~M.}\ \bibnamefont {Bakkers}}, \ and\ \bibinfo
  {author} {\bibfnamefont {L.~P.}\ \bibnamefont {Kouwenhoven}},\ }\bibfield
  {title} {\enquote {\bibinfo {title} {Signatures of {Majorana} fermions in
  hybrid superconductor-semiconductor nanowire devices},}\ }\href {\doibase
  10.1126/science.1222360} {\bibfield  {journal} {\bibinfo  {journal}
  {Science}\ }\textbf {\bibinfo {volume} {336}},\ \bibinfo {pages} {1003--1007}
  (\bibinfo {year} {2012})},\ \Eprint
  {http://arxiv.org/abs/https://science.sciencemag.org/content/336/6084/1003.full.pdf}
  {https://science.sciencemag.org/content/336/6084/1003.full.pdf} \BibitemShut
  {NoStop}%
\bibitem [{\citenamefont {Albrecht}\ \emph {et~al.}(2016)\citenamefont
  {Albrecht}, \citenamefont {Higginbotham}, \citenamefont {Madsen},
  \citenamefont {Kuemmeth}, \citenamefont {Jespersen}, \citenamefont
  {Nyg{\aa}rd}, \citenamefont {Krogstrup},\ and\ \citenamefont
  {Marcus}}]{AlbrechtNat2016}%
  \BibitemOpen
  \bibfield  {author} {\bibinfo {author} {\bibfnamefont {S.~M.}\ \bibnamefont
  {Albrecht}}, \bibinfo {author} {\bibfnamefont {A.~P.}\ \bibnamefont
  {Higginbotham}}, \bibinfo {author} {\bibfnamefont {M.}~\bibnamefont
  {Madsen}}, \bibinfo {author} {\bibfnamefont {F.}~\bibnamefont {Kuemmeth}},
  \bibinfo {author} {\bibfnamefont {T.~S.}\ \bibnamefont {Jespersen}}, \bibinfo
  {author} {\bibfnamefont {J.}~\bibnamefont {Nyg{\aa}rd}}, \bibinfo {author}
  {\bibfnamefont {P.}~\bibnamefont {Krogstrup}}, \ and\ \bibinfo {author}
  {\bibfnamefont {C.~M.}\ \bibnamefont {Marcus}},\ }\bibfield  {title}
  {\enquote {\bibinfo {title} {Exponential protection of zero modes in
  {Majorana} islands},}\ }\href {\doibase 10.1038/nature17162} {\bibfield
  {journal} {\bibinfo  {journal} {Nature}\ }\textbf {\bibinfo {volume} {531}},\
  \bibinfo {pages} {206--209} (\bibinfo {year} {2016})}\BibitemShut {NoStop}%
\bibitem [{\citenamefont {Alicea}(2012)}]{AliceaRPP2012}%
  \BibitemOpen
  \bibfield  {author} {\bibinfo {author} {\bibfnamefont {Jason}\ \bibnamefont
  {Alicea}},\ }\bibfield  {title} {\enquote {\bibinfo {title} {New directions
  in the pursuit of {Majorana} fermions in solid state systems},}\ }\href
  {\doibase 10.1088/0034-4885/75/7/076501} {\bibfield  {journal} {\bibinfo
  {journal} {Reports on Progress in Physics}\ }\textbf {\bibinfo {volume}
  {75}},\ \bibinfo {pages} {076501} (\bibinfo {year} {2012})}\BibitemShut
  {NoStop}%
\bibitem [{\citenamefont {Beenakker}(2015)}]{BeenakkerRMP2015}%
  \BibitemOpen
  \bibfield  {author} {\bibinfo {author} {\bibfnamefont {C.~W.~J.}\
  \bibnamefont {Beenakker}},\ }\bibfield  {title} {\enquote {\bibinfo {title}
  {Random-matrix theory of {Majorana} fermions and topological
  superconductors},}\ }\href {\doibase 10.1103/RevModPhys.87.1037} {\bibfield
  {journal} {\bibinfo  {journal} {Rev. Mod. Phys.}\ }\textbf {\bibinfo {volume}
  {87}},\ \bibinfo {pages} {1037--1066} (\bibinfo {year} {2015})}\BibitemShut
  {NoStop}%
\bibitem [{\citenamefont {Sarma}\ \emph {et~al.}(2015)\citenamefont {Sarma},
  \citenamefont {Freedman},\ and\ \citenamefont {Nayak}}]{DasSarmaNPJQI2015}%
  \BibitemOpen
  \bibfield  {author} {\bibinfo {author} {\bibfnamefont {Sankar~Das}\
  \bibnamefont {Sarma}}, \bibinfo {author} {\bibfnamefont {Michael}\
  \bibnamefont {Freedman}}, \ and\ \bibinfo {author} {\bibfnamefont {Chetan}\
  \bibnamefont {Nayak}},\ }\bibfield  {title} {\enquote {\bibinfo {title}
  {Majorana zero modes and topological quantum computation},}\ }\href {\doibase
  10.1038/npjqi.2015.1} {\bibfield  {journal} {\bibinfo  {journal} {npj Quantum
  Information}\ }\textbf {\bibinfo {volume} {1}},\ \bibinfo {pages} {15001}
  (\bibinfo {year} {2015})}\BibitemShut {NoStop}%
\bibitem [{\citenamefont {Beenakker}\ and\ \citenamefont
  {Kouwenhoven}(2016)}]{BeenakkerNatPhys2016}%
  \BibitemOpen
  \bibfield  {author} {\bibinfo {author} {\bibfnamefont {Carlo}\ \bibnamefont
  {Beenakker}}\ and\ \bibinfo {author} {\bibfnamefont {Leo}\ \bibnamefont
  {Kouwenhoven}},\ }\bibfield  {title} {\enquote {\bibinfo {title} {A road to
  reality with topological superconductors},}\ }\href {\doibase
  10.1038/nphys3778} {\bibfield  {journal} {\bibinfo  {journal} {Nature
  Physics}\ }\textbf {\bibinfo {volume} {12}},\ \bibinfo {pages} {618--621}
  (\bibinfo {year} {2016})}\BibitemShut {NoStop}%
\bibitem [{\citenamefont {Aguado}(2017)}]{AguadoReview}%
  \BibitemOpen
  \bibfield  {author} {\bibinfo {author} {\bibfnamefont {Ramon}\ \bibnamefont
  {Aguado}},\ }\bibfield  {title} {\enquote {\bibinfo {title} {{Majorana
  quasiparticles in condensed matter}},}\ }\href {\doibase
  10.1393/ncr/i2017-10141-9} {\bibfield  {journal} {\bibinfo  {journal} {Riv.
  Nuomo. Cim.}\ }\textbf {\bibinfo {volume} {40}} (\bibinfo {year} {2017}),\
  10.1393/ncr/i2017-10141-9}\BibitemShut {NoStop}%
\bibitem [{\citenamefont {Lutchyn}\ \emph {et~al.}(2018)\citenamefont
  {Lutchyn}, \citenamefont {Bakkers}, \citenamefont {Kouwenhoven},
  \citenamefont {Krogstrup}, \citenamefont {Marcus},\ and\ \citenamefont
  {Oreg}}]{LutchynNatRevMat2018}%
  \BibitemOpen
  \bibfield  {author} {\bibinfo {author} {\bibfnamefont {R.~M.}\ \bibnamefont
  {Lutchyn}}, \bibinfo {author} {\bibfnamefont {E.~P. A.~M.}\ \bibnamefont
  {Bakkers}}, \bibinfo {author} {\bibfnamefont {L.~P.}\ \bibnamefont
  {Kouwenhoven}}, \bibinfo {author} {\bibfnamefont {P.}~\bibnamefont
  {Krogstrup}}, \bibinfo {author} {\bibfnamefont {C.~M.}\ \bibnamefont
  {Marcus}}, \ and\ \bibinfo {author} {\bibfnamefont {Y.}~\bibnamefont
  {Oreg}},\ }\bibfield  {title} {\enquote {\bibinfo {title} {Majorana zero
  modes in superconductor--semiconductor heterostructures},}\ }\href {\doibase
  10.1038/s41578-018-0003-1} {\bibfield  {journal} {\bibinfo  {journal} {Nature
  Reviews Materials}\ }\textbf {\bibinfo {volume} {3}},\ \bibinfo {pages}
  {52--68} (\bibinfo {year} {2018})}\BibitemShut {NoStop}%
\bibitem [{\citenamefont {Beenakker}(2020)}]{BeenakkerSciPost2020}%
  \BibitemOpen
  \bibfield  {author} {\bibinfo {author} {\bibfnamefont {C.~W.~J.}\
  \bibnamefont {Beenakker}},\ }\bibfield  {title} {\enquote {\bibinfo {title}
  {{Search for non-Abelian Majorana braiding statistics in superconductors}},}\
  }\href {\doibase 10.21468/SciPostPhysLectNotes.15} {\bibfield  {journal}
  {\bibinfo  {journal} {SciPost Phys. Lect. Notes}\ ,\ \bibinfo {pages} {15}}
  (\bibinfo {year} {2020})}\BibitemShut {NoStop}%
\bibitem [{\citenamefont {Prada}\ \emph {et~al.}(2020)\citenamefont {Prada},
  \citenamefont {San-Jose}, \citenamefont {de~Moor}, \citenamefont {Geresdi},
  \citenamefont {Lee}, \citenamefont {Klinovaja}, \citenamefont {Loss},
  \citenamefont {Nyg{\aa}rd}, \citenamefont {Aguado},\ and\ \citenamefont
  {Kouwenhoven}}]{PradaReview}%
  \BibitemOpen
  \bibfield  {author} {\bibinfo {author} {\bibfnamefont {Elsa}\ \bibnamefont
  {Prada}}, \bibinfo {author} {\bibfnamefont {Pablo}\ \bibnamefont {San-Jose}},
  \bibinfo {author} {\bibfnamefont {Michiel W.~A.}\ \bibnamefont {de~Moor}},
  \bibinfo {author} {\bibfnamefont {Attila}\ \bibnamefont {Geresdi}}, \bibinfo
  {author} {\bibfnamefont {Eduardo J.~H.}\ \bibnamefont {Lee}}, \bibinfo
  {author} {\bibfnamefont {Jelena}\ \bibnamefont {Klinovaja}}, \bibinfo
  {author} {\bibfnamefont {Daniel}\ \bibnamefont {Loss}}, \bibinfo {author}
  {\bibfnamefont {Jesper}\ \bibnamefont {Nyg{\aa}rd}}, \bibinfo {author}
  {\bibfnamefont {Ram{\'o}n}\ \bibnamefont {Aguado}}, \ and\ \bibinfo {author}
  {\bibfnamefont {Leo~P.}\ \bibnamefont {Kouwenhoven}},\ }\bibfield  {title}
  {\enquote {\bibinfo {title} {From {Andreev} to {Majorana} bound states in
  hybrid superconductor--semiconductor nanowires},}\ }\href@noop {} {\bibfield
  {journal} {\bibinfo  {journal} {Nature Reviews Physics}\ }\textbf {\bibinfo
  {volume} {2}},\ \bibinfo {pages} {575--594} (\bibinfo {year}
  {2020})}\BibitemShut {NoStop}%
\bibitem [{\citenamefont {Scheurer}\ and\ \citenamefont
  {Shnirman}(2013)}]{ScheurerPRB2013}%
  \BibitemOpen
  \bibfield  {author} {\bibinfo {author} {\bibfnamefont {M.~S.}\ \bibnamefont
  {Scheurer}}\ and\ \bibinfo {author} {\bibfnamefont {A.}~\bibnamefont
  {Shnirman}},\ }\bibfield  {title} {\enquote {\bibinfo {title} {Nonadiabatic
  processes in {Majorana} qubit systems},}\ }\href {\doibase
  10.1103/PhysRevB.88.064515} {\bibfield  {journal} {\bibinfo  {journal} {Phys.
  Rev. B}\ }\textbf {\bibinfo {volume} {88}},\ \bibinfo {pages} {064515}
  (\bibinfo {year} {2013})}\BibitemShut {NoStop}%
\bibitem [{\citenamefont {Aseev}\ \emph {et~al.}(2018)\citenamefont {Aseev},
  \citenamefont {Klinovaja},\ and\ \citenamefont {Loss}}]{AseevPRB2018}%
  \BibitemOpen
  \bibfield  {author} {\bibinfo {author} {\bibfnamefont {Pavel~P.}\
  \bibnamefont {Aseev}}, \bibinfo {author} {\bibfnamefont {Jelena}\
  \bibnamefont {Klinovaja}}, \ and\ \bibinfo {author} {\bibfnamefont {Daniel}\
  \bibnamefont {Loss}},\ }\bibfield  {title} {\enquote {\bibinfo {title}
  {Lifetime of {Majorana} qubits in {Rashba} nanowires with nonuniform chemical
  potential},}\ }\href {\doibase 10.1103/PhysRevB.98.155414} {\bibfield
  {journal} {\bibinfo  {journal} {Phys. Rev. B}\ }\textbf {\bibinfo {volume}
  {98}},\ \bibinfo {pages} {155414} (\bibinfo {year} {2018})}\BibitemShut
  {NoStop}%
\bibitem [{\citenamefont {Aseev}\ \emph {et~al.}(2019)\citenamefont {Aseev},
  \citenamefont {Marra}, \citenamefont {Stano}, \citenamefont {Klinovaja},\
  and\ \citenamefont {Loss}}]{AseevPRB2019}%
  \BibitemOpen
  \bibfield  {author} {\bibinfo {author} {\bibfnamefont {Pavel~P.}\
  \bibnamefont {Aseev}}, \bibinfo {author} {\bibfnamefont {Pasquale}\
  \bibnamefont {Marra}}, \bibinfo {author} {\bibfnamefont {Peter}\ \bibnamefont
  {Stano}}, \bibinfo {author} {\bibfnamefont {Jelena}\ \bibnamefont
  {Klinovaja}}, \ and\ \bibinfo {author} {\bibfnamefont {Daniel}\ \bibnamefont
  {Loss}},\ }\bibfield  {title} {\enquote {\bibinfo {title} {Degeneracy lifting
  of {Majorana} bound states due to electron-phonon interactions},}\ }\href
  {\doibase 10.1103/PhysRevB.99.205435} {\bibfield  {journal} {\bibinfo
  {journal} {Phys. Rev. B}\ }\textbf {\bibinfo {volume} {99}},\ \bibinfo
  {pages} {205435} (\bibinfo {year} {2019})}\BibitemShut {NoStop}%
\bibitem [{\citenamefont {Budich}\ \emph
  {et~al.}(2012{\natexlab{a}})\citenamefont {Budich}, \citenamefont {Walter},\
  and\ \citenamefont {Trauzettel}}]{WalterPRB2012}%
  \BibitemOpen
  \bibfield  {author} {\bibinfo {author} {\bibfnamefont {Jan~Carl}\
  \bibnamefont {Budich}}, \bibinfo {author} {\bibfnamefont {Stefan}\
  \bibnamefont {Walter}}, \ and\ \bibinfo {author} {\bibfnamefont {Bj\"orn}\
  \bibnamefont {Trauzettel}},\ }\bibfield  {title} {\enquote {\bibinfo {title}
  {Failure of protection of {Majorana} based qubits against decoherence},}\
  }\href {\doibase 10.1103/PhysRevB.85.121405} {\bibfield  {journal} {\bibinfo
  {journal} {Phys. Rev. B}\ }\textbf {\bibinfo {volume} {85}},\ \bibinfo
  {pages} {121405} (\bibinfo {year} {2012}{\natexlab{a}})}\BibitemShut
  {NoStop}%
\bibitem [{\citenamefont {Knapp}\ \emph {et~al.}(2018)\citenamefont {Knapp},
  \citenamefont {Karzig}, \citenamefont {Lutchyn},\ and\ \citenamefont
  {Nayak}}]{KnappPRB2018}%
  \BibitemOpen
  \bibfield  {author} {\bibinfo {author} {\bibfnamefont {Christina}\
  \bibnamefont {Knapp}}, \bibinfo {author} {\bibfnamefont {Torsten}\
  \bibnamefont {Karzig}}, \bibinfo {author} {\bibfnamefont {Roman~M.}\
  \bibnamefont {Lutchyn}}, \ and\ \bibinfo {author} {\bibfnamefont {Chetan}\
  \bibnamefont {Nayak}},\ }\bibfield  {title} {\enquote {\bibinfo {title}
  {Dephasing of {Majorana}-based qubits},}\ }\href {\doibase
  10.1103/PhysRevB.97.125404} {\bibfield  {journal} {\bibinfo  {journal} {Phys.
  Rev. B}\ }\textbf {\bibinfo {volume} {97}},\ \bibinfo {pages} {125404}
  (\bibinfo {year} {2018})}\BibitemShut {NoStop}%
\bibitem [{\citenamefont {Aasen}\ \emph {et~al.}(2016)\citenamefont {Aasen},
  \citenamefont {Hell}, \citenamefont {Mishmash}, \citenamefont {Higginbotham},
  \citenamefont {Danon}, \citenamefont {Leijnse}, \citenamefont {Jespersen},
  \citenamefont {Folk}, \citenamefont {Marcus}, \citenamefont {Flensberg},\
  and\ \citenamefont {Alicea}}]{AasenPRX2016}%
  \BibitemOpen
  \bibfield  {author} {\bibinfo {author} {\bibfnamefont {David}\ \bibnamefont
  {Aasen}}, \bibinfo {author} {\bibfnamefont {Michael}\ \bibnamefont {Hell}},
  \bibinfo {author} {\bibfnamefont {Ryan~V.}\ \bibnamefont {Mishmash}},
  \bibinfo {author} {\bibfnamefont {Andrew}\ \bibnamefont {Higginbotham}},
  \bibinfo {author} {\bibfnamefont {Jeroen}\ \bibnamefont {Danon}}, \bibinfo
  {author} {\bibfnamefont {Martin}\ \bibnamefont {Leijnse}}, \bibinfo {author}
  {\bibfnamefont {Thomas~S.}\ \bibnamefont {Jespersen}}, \bibinfo {author}
  {\bibfnamefont {Joshua~A.}\ \bibnamefont {Folk}}, \bibinfo {author}
  {\bibfnamefont {Charles~M.}\ \bibnamefont {Marcus}}, \bibinfo {author}
  {\bibfnamefont {Karsten}\ \bibnamefont {Flensberg}}, \ and\ \bibinfo {author}
  {\bibfnamefont {Jason}\ \bibnamefont {Alicea}},\ }\bibfield  {title}
  {\enquote {\bibinfo {title} {Milestones toward {Majorana}-based quantum
  computing},}\ }\href {\doibase 10.1103/PhysRevX.6.031016} {\bibfield
  {journal} {\bibinfo  {journal} {Phys. Rev. X}\ }\textbf {\bibinfo {volume}
  {6}},\ \bibinfo {pages} {031016} (\bibinfo {year} {2016})}\BibitemShut
  {NoStop}%
\bibitem [{\citenamefont {Karzig}\ \emph {et~al.}(2017)\citenamefont {Karzig},
  \citenamefont {Knapp}, \citenamefont {Lutchyn}, \citenamefont {Bonderson},
  \citenamefont {Hastings}, \citenamefont {Nayak}, \citenamefont {Alicea},
  \citenamefont {Flensberg}, \citenamefont {Plugge}, \citenamefont {Oreg},
  \citenamefont {Marcus},\ and\ \citenamefont {Freedman}}]{KarzigPRB2017}%
  \BibitemOpen
  \bibfield  {author} {\bibinfo {author} {\bibfnamefont {Torsten}\ \bibnamefont
  {Karzig}}, \bibinfo {author} {\bibfnamefont {Christina}\ \bibnamefont
  {Knapp}}, \bibinfo {author} {\bibfnamefont {Roman~M.}\ \bibnamefont
  {Lutchyn}}, \bibinfo {author} {\bibfnamefont {Parsa}\ \bibnamefont
  {Bonderson}}, \bibinfo {author} {\bibfnamefont {Matthew~B.}\ \bibnamefont
  {Hastings}}, \bibinfo {author} {\bibfnamefont {Chetan}\ \bibnamefont
  {Nayak}}, \bibinfo {author} {\bibfnamefont {Jason}\ \bibnamefont {Alicea}},
  \bibinfo {author} {\bibfnamefont {Karsten}\ \bibnamefont {Flensberg}},
  \bibinfo {author} {\bibfnamefont {Stephan}\ \bibnamefont {Plugge}}, \bibinfo
  {author} {\bibfnamefont {Yuval}\ \bibnamefont {Oreg}}, \bibinfo {author}
  {\bibfnamefont {Charles~M.}\ \bibnamefont {Marcus}}, \ and\ \bibinfo {author}
  {\bibfnamefont {Michael~H.}\ \bibnamefont {Freedman}},\ }\bibfield  {title}
  {\enquote {\bibinfo {title} {Scalable designs for
  quasiparticle-poisoning-protected topological quantum computation with
  {Majorana} zero modes},}\ }\href {\doibase 10.1103/PhysRevB.95.235305}
  {\bibfield  {journal} {\bibinfo  {journal} {Phys. Rev. B}\ }\textbf {\bibinfo
  {volume} {95}},\ \bibinfo {pages} {235305} (\bibinfo {year}
  {2017})}\BibitemShut {NoStop}%
\bibitem [{\citenamefont {Hell}\ \emph {et~al.}(2016)\citenamefont {Hell},
  \citenamefont {Danon}, \citenamefont {Flensberg},\ and\ \citenamefont
  {Leijnse}}]{HellPRB2016}%
  \BibitemOpen
  \bibfield  {author} {\bibinfo {author} {\bibfnamefont {Michael}\ \bibnamefont
  {Hell}}, \bibinfo {author} {\bibfnamefont {Jeroen}\ \bibnamefont {Danon}},
  \bibinfo {author} {\bibfnamefont {Karsten}\ \bibnamefont {Flensberg}}, \ and\
  \bibinfo {author} {\bibfnamefont {Martin}\ \bibnamefont {Leijnse}},\
  }\bibfield  {title} {\enquote {\bibinfo {title} {Time scales for {Majorana}
  manipulation using {Coulomb} blockade in gate-controlled superconducting
  nanowires},}\ }\href {\doibase 10.1103/PhysRevB.94.035424} {\bibfield
  {journal} {\bibinfo  {journal} {Phys. Rev. B}\ }\textbf {\bibinfo {volume}
  {94}},\ \bibinfo {pages} {035424} (\bibinfo {year} {2016})}\BibitemShut
  {NoStop}%
\bibitem [{\citenamefont {Rahmani}\ \emph {et~al.}(2017)\citenamefont
  {Rahmani}, \citenamefont {Seradjeh},\ and\ \citenamefont
  {Franz}}]{RahmaniPRB2017}%
  \BibitemOpen
  \bibfield  {author} {\bibinfo {author} {\bibfnamefont {Armin}\ \bibnamefont
  {Rahmani}}, \bibinfo {author} {\bibfnamefont {Babak}\ \bibnamefont
  {Seradjeh}}, \ and\ \bibinfo {author} {\bibfnamefont {Marcel}\ \bibnamefont
  {Franz}},\ }\bibfield  {title} {\enquote {\bibinfo {title} {Optimal diabatic
  dynamics of {Majorana}-based quantum gates},}\ }\href {\doibase
  10.1103/PhysRevB.96.075158} {\bibfield  {journal} {\bibinfo  {journal} {Phys.
  Rev. B}\ }\textbf {\bibinfo {volume} {96}},\ \bibinfo {pages} {075158}
  (\bibinfo {year} {2017})}\BibitemShut {NoStop}%
\bibitem [{\citenamefont {Breckwoldt}\ \emph {et~al.}(2022)\citenamefont
  {Breckwoldt}, \citenamefont {Posske},\ and\ \citenamefont
  {Thorwart}}]{Breckwoldt_2022}%
  \BibitemOpen
  \bibfield  {author} {\bibinfo {author} {\bibfnamefont {Niels}\ \bibnamefont
  {Breckwoldt}}, \bibinfo {author} {\bibfnamefont {Thore}\ \bibnamefont
  {Posske}}, \ and\ \bibinfo {author} {\bibfnamefont {Michael}\ \bibnamefont
  {Thorwart}},\ }\bibfield  {title} {\enquote {\bibinfo {title} {Bath-induced
  decoherence in finite-size {Majorana} wires at non-zero temperature},}\
  }\href {\doibase 10.1088/1367-2630/ac46e2} {\bibfield  {journal} {\bibinfo
  {journal} {New Journal of Physics}\ }\textbf {\bibinfo {volume} {24}},\
  \bibinfo {pages} {013033} (\bibinfo {year} {2022})}\BibitemShut {NoStop}%
\bibitem [{\citenamefont {Bauer}\ \emph {et~al.}(2018)\citenamefont {Bauer},
  \citenamefont {Karzig}, \citenamefont {Mishmash}, \citenamefont {Antipov},\
  and\ \citenamefont {Alicea}}]{BauerSciPost2018}%
  \BibitemOpen
  \bibfield  {author} {\bibinfo {author} {\bibfnamefont {Bela}\ \bibnamefont
  {Bauer}}, \bibinfo {author} {\bibfnamefont {Torsten}\ \bibnamefont {Karzig}},
  \bibinfo {author} {\bibfnamefont {Ryan~V.}\ \bibnamefont {Mishmash}},
  \bibinfo {author} {\bibfnamefont {Andrey~E.}\ \bibnamefont {Antipov}}, \ and\
  \bibinfo {author} {\bibfnamefont {Jason}\ \bibnamefont {Alicea}},\ }\bibfield
   {title} {\enquote {\bibinfo {title} {{Dynamics of {Majorana}-based qubits
  operated with an array of tunable gates}},}\ }\href {\doibase
  10.21468/SciPostPhys.5.1.004} {\bibfield  {journal} {\bibinfo  {journal}
  {SciPost Phys.}\ }\textbf {\bibinfo {volume} {5}},\ \bibinfo {pages} {4}
  (\bibinfo {year} {2018})}\BibitemShut {NoStop}%
\bibitem [{\citenamefont {Fulga}\ \emph
  {et~al.}(2013{\natexlab{a}})\citenamefont {Fulga}, \citenamefont {van Heck},
  \citenamefont {Burrello},\ and\ \citenamefont {Hyart}}]{FulgaPRB2013}%
  \BibitemOpen
  \bibfield  {author} {\bibinfo {author} {\bibfnamefont {I.~C.}\ \bibnamefont
  {Fulga}}, \bibinfo {author} {\bibfnamefont {B.}~\bibnamefont {van Heck}},
  \bibinfo {author} {\bibfnamefont {M.}~\bibnamefont {Burrello}}, \ and\
  \bibinfo {author} {\bibfnamefont {T.}~\bibnamefont {Hyart}},\ }\bibfield
  {title} {\enquote {\bibinfo {title} {Effects of disorder on
  {Coulomb}-assisted braiding of {Majorana} zero modes},}\ }\href {\doibase
  10.1103/PhysRevB.88.155435} {\bibfield  {journal} {\bibinfo  {journal} {Phys.
  Rev. B}\ }\textbf {\bibinfo {volume} {88}},\ \bibinfo {pages} {155435}
  (\bibinfo {year} {2013}{\natexlab{a}})}\BibitemShut {NoStop}%
\bibitem [{\citenamefont {Zhang}\ \emph {et~al.}(2019)\citenamefont {Zhang},
  \citenamefont {Mei}, \citenamefont {Meng}, \citenamefont {Liang},\ and\
  \citenamefont {Yang}}]{ZhangPRA2019}%
  \BibitemOpen
  \bibfield  {author} {\bibinfo {author} {\bibfnamefont {Zhen-Tao}\
  \bibnamefont {Zhang}}, \bibinfo {author} {\bibfnamefont {Feng}\ \bibnamefont
  {Mei}}, \bibinfo {author} {\bibfnamefont {Xiang-Guo}\ \bibnamefont {Meng}},
  \bibinfo {author} {\bibfnamefont {Bao-Long}\ \bibnamefont {Liang}}, \ and\
  \bibinfo {author} {\bibfnamefont {Zhen-Shan}\ \bibnamefont {Yang}},\
  }\bibfield  {title} {\enquote {\bibinfo {title} {Effects of decoherence on
  diabatic errors in {Majorana} braiding},}\ }\href {\doibase
  10.1103/PhysRevA.100.012324} {\bibfield  {journal} {\bibinfo  {journal}
  {Phys. Rev. A}\ }\textbf {\bibinfo {volume} {100}},\ \bibinfo {pages}
  {012324} (\bibinfo {year} {2019})}\BibitemShut {NoStop}%
\bibitem [{\citenamefont {Kornich}\ \emph {et~al.}(2021)\citenamefont
  {Kornich}, \citenamefont {Huang}, \citenamefont {Repin},\ and\ \citenamefont
  {Nazarov}}]{KornichPRL2021}%
  \BibitemOpen
  \bibfield  {author} {\bibinfo {author} {\bibfnamefont {Viktoriia}\
  \bibnamefont {Kornich}}, \bibinfo {author} {\bibfnamefont {Xiaoli}\
  \bibnamefont {Huang}}, \bibinfo {author} {\bibfnamefont {Evgeny}\
  \bibnamefont {Repin}}, \ and\ \bibinfo {author} {\bibfnamefont {Yuli~V.}\
  \bibnamefont {Nazarov}},\ }\bibfield  {title} {\enquote {\bibinfo {title}
  {Braiding and all quantum operations with {Majorana} modes in {1D}},}\ }\href
  {\doibase 10.1103/PhysRevLett.126.117701} {\bibfield  {journal} {\bibinfo
  {journal} {Phys. Rev. Lett.}\ }\textbf {\bibinfo {volume} {126}},\ \bibinfo
  {pages} {117701} (\bibinfo {year} {2021})}\BibitemShut {NoStop}%
\bibitem [{\citenamefont {Boross}\ and\ \citenamefont
  {P\'alyi}(2022)}]{BorossPRB2022}%
  \BibitemOpen
  \bibfield  {author} {\bibinfo {author} {\bibfnamefont {P\'eter}\ \bibnamefont
  {Boross}}\ and\ \bibinfo {author} {\bibfnamefont {Andr\'as}\ \bibnamefont
  {P\'alyi}},\ }\bibfield  {title} {\enquote {\bibinfo {title} {Dephasing of
  {Majorana} qubits due to quasistatic disorder},}\ }\href {\doibase
  10.1103/PhysRevB.105.035413} {\bibfield  {journal} {\bibinfo  {journal}
  {Phys. Rev. B}\ }\textbf {\bibinfo {volume} {105}},\ \bibinfo {pages}
  {035413} (\bibinfo {year} {2022})}\BibitemShut {NoStop}%
\bibitem [{\citenamefont {Brouwer}\ \emph {et~al.}(2011)\citenamefont
  {Brouwer}, \citenamefont {Duckheim}, \citenamefont {Romito},\ and\
  \citenamefont {von Oppen}}]{BrouwerPRL2011}%
  \BibitemOpen
  \bibfield  {author} {\bibinfo {author} {\bibfnamefont {Piet~W.}\ \bibnamefont
  {Brouwer}}, \bibinfo {author} {\bibfnamefont {Mathias}\ \bibnamefont
  {Duckheim}}, \bibinfo {author} {\bibfnamefont {Alessandro}\ \bibnamefont
  {Romito}}, \ and\ \bibinfo {author} {\bibfnamefont {Felix}\ \bibnamefont {von
  Oppen}},\ }\bibfield  {title} {\enquote {\bibinfo {title} {Probability
  distribution of {Majorana} end-state energies in disordered wires},}\ }\href
  {\doibase 10.1103/PhysRevLett.107.196804} {\bibfield  {journal} {\bibinfo
  {journal} {Phys. Rev. Lett.}\ }\textbf {\bibinfo {volume} {107}},\ \bibinfo
  {pages} {196804} (\bibinfo {year} {2011})}\BibitemShut {NoStop}%
\bibitem [{\citenamefont {Goldstein}\ and\ \citenamefont
  {Chamon}(2011)}]{GoldsteinPRB2011}%
  \BibitemOpen
  \bibfield  {author} {\bibinfo {author} {\bibfnamefont {G.}~\bibnamefont
  {Goldstein}}\ and\ \bibinfo {author} {\bibfnamefont {C.}~\bibnamefont
  {Chamon}},\ }\bibfield  {title} {\enquote {\bibinfo {title} {Decay rates for
  topological memories encoded with {Majorana} fermions},}\ }\href {\doibase
  10.1103/PhysRevB.84.205109} {\bibfield  {journal} {\bibinfo  {journal} {Phys.
  Rev. B}\ }\textbf {\bibinfo {volume} {84}},\ \bibinfo {pages} {205109}
  (\bibinfo {year} {2011})}\BibitemShut {NoStop}%
\bibitem [{\citenamefont {Schmidt}\ \emph {et~al.}(2012)\citenamefont
  {Schmidt}, \citenamefont {Rainis},\ and\ \citenamefont
  {Loss}}]{SchmidtPRB2012}%
  \BibitemOpen
  \bibfield  {author} {\bibinfo {author} {\bibfnamefont {Manuel~J.}\
  \bibnamefont {Schmidt}}, \bibinfo {author} {\bibfnamefont {Diego}\
  \bibnamefont {Rainis}}, \ and\ \bibinfo {author} {\bibfnamefont {Daniel}\
  \bibnamefont {Loss}},\ }\bibfield  {title} {\enquote {\bibinfo {title}
  {Decoherence of {Majorana} qubits by noisy gates},}\ }\href {\doibase
  10.1103/PhysRevB.86.085414} {\bibfield  {journal} {\bibinfo  {journal} {Phys.
  Rev. B}\ }\textbf {\bibinfo {volume} {86}},\ \bibinfo {pages} {085414}
  (\bibinfo {year} {2012})}\BibitemShut {NoStop}%
\bibitem [{\citenamefont {Budich}\ \emph
  {et~al.}(2012{\natexlab{b}})\citenamefont {Budich}, \citenamefont {Walter},\
  and\ \citenamefont {Trauzettel}}]{BudichPRB2012}%
  \BibitemOpen
  \bibfield  {author} {\bibinfo {author} {\bibfnamefont {Jan~Carl}\
  \bibnamefont {Budich}}, \bibinfo {author} {\bibfnamefont {Stefan}\
  \bibnamefont {Walter}}, \ and\ \bibinfo {author} {\bibfnamefont {Bj\"orn}\
  \bibnamefont {Trauzettel}},\ }\bibfield  {title} {\enquote {\bibinfo {title}
  {Failure of protection of {Majorana} based qubits against decoherence},}\
  }\href {\doibase 10.1103/PhysRevB.85.121405} {\bibfield  {journal} {\bibinfo
  {journal} {Phys. Rev. B}\ }\textbf {\bibinfo {volume} {85}},\ \bibinfo
  {pages} {121405} (\bibinfo {year} {2012}{\natexlab{b}})}\BibitemShut
  {NoStop}%
\bibitem [{\citenamefont {Rainis}\ and\ \citenamefont
  {Loss}(2012)}]{RainisPRB2012}%
  \BibitemOpen
  \bibfield  {author} {\bibinfo {author} {\bibfnamefont {Diego}\ \bibnamefont
  {Rainis}}\ and\ \bibinfo {author} {\bibfnamefont {Daniel}\ \bibnamefont
  {Loss}},\ }\bibfield  {title} {\enquote {\bibinfo {title} {Majorana qubit
  decoherence by quasiparticle poisoning},}\ }\href {\doibase
  10.1103/PhysRevB.85.174533} {\bibfield  {journal} {\bibinfo  {journal} {Phys.
  Rev. B}\ }\textbf {\bibinfo {volume} {85}},\ \bibinfo {pages} {174533}
  (\bibinfo {year} {2012})}\BibitemShut {NoStop}%
\bibitem [{\citenamefont {Pedrocchi}\ and\ \citenamefont
  {DiVincenzo}(2015)}]{PedrocchiPRL2015}%
  \BibitemOpen
  \bibfield  {author} {\bibinfo {author} {\bibfnamefont {Fabio~L.}\
  \bibnamefont {Pedrocchi}}\ and\ \bibinfo {author} {\bibfnamefont {David~P.}\
  \bibnamefont {DiVincenzo}},\ }\bibfield  {title} {\enquote {\bibinfo {title}
  {Majorana braiding with thermal noise},}\ }\href {\doibase
  10.1103/PhysRevLett.115.120402} {\bibfield  {journal} {\bibinfo  {journal}
  {Phys. Rev. Lett.}\ }\textbf {\bibinfo {volume} {115}},\ \bibinfo {pages}
  {120402} (\bibinfo {year} {2015})}\BibitemShut {NoStop}%
\bibitem [{\citenamefont {Lai}\ \emph {et~al.}(2018)\citenamefont {Lai},
  \citenamefont {Yang}, \citenamefont {Huang},\ and\ \citenamefont
  {Zhang}}]{LaiPRB2018}%
  \BibitemOpen
  \bibfield  {author} {\bibinfo {author} {\bibfnamefont {Hon-Lam}\ \bibnamefont
  {Lai}}, \bibinfo {author} {\bibfnamefont {Pei-Yun}\ \bibnamefont {Yang}},
  \bibinfo {author} {\bibfnamefont {Yu-Wei}\ \bibnamefont {Huang}}, \ and\
  \bibinfo {author} {\bibfnamefont {Wei-Min}\ \bibnamefont {Zhang}},\
  }\bibfield  {title} {\enquote {\bibinfo {title} {Exact master equation and
  non-{Markovian} decoherence dynamics of {Majorana} zero modes under
  gate-induced charge fluctuations},}\ }\href {\doibase
  10.1103/PhysRevB.97.054508} {\bibfield  {journal} {\bibinfo  {journal} {Phys.
  Rev. B}\ }\textbf {\bibinfo {volume} {97}},\ \bibinfo {pages} {054508}
  (\bibinfo {year} {2018})}\BibitemShut {NoStop}%
\bibitem [{\citenamefont {Mishmash}\ \emph {et~al.}(2020)\citenamefont
  {Mishmash}, \citenamefont {Bauer}, \citenamefont {von Oppen},\ and\
  \citenamefont {Alicea}}]{MishmashPRB2020}%
  \BibitemOpen
  \bibfield  {author} {\bibinfo {author} {\bibfnamefont {Ryan~V.}\ \bibnamefont
  {Mishmash}}, \bibinfo {author} {\bibfnamefont {Bela}\ \bibnamefont {Bauer}},
  \bibinfo {author} {\bibfnamefont {Felix}\ \bibnamefont {von Oppen}}, \ and\
  \bibinfo {author} {\bibfnamefont {Jason}\ \bibnamefont {Alicea}},\ }\bibfield
   {title} {\enquote {\bibinfo {title} {Dephasing and leakage dynamics of noisy
  {Majorana}-based qubits: Topological versus {Andreev}},}\ }\href {\doibase
  10.1103/PhysRevB.101.075404} {\bibfield  {journal} {\bibinfo  {journal}
  {Phys. Rev. B}\ }\textbf {\bibinfo {volume} {101}},\ \bibinfo {pages}
  {075404} (\bibinfo {year} {2020})}\BibitemShut {NoStop}%
\bibitem [{\citenamefont {Tutschku}\ \emph {et~al.}(2020)\citenamefont
  {Tutschku}, \citenamefont {Reinthaler}, \citenamefont {Lei}, \citenamefont
  {MacDonald},\ and\ \citenamefont {Hankiewicz}}]{TutschkuPRB2020}%
  \BibitemOpen
  \bibfield  {author} {\bibinfo {author} {\bibfnamefont {C.}~\bibnamefont
  {Tutschku}}, \bibinfo {author} {\bibfnamefont {R.~W.}\ \bibnamefont
  {Reinthaler}}, \bibinfo {author} {\bibfnamefont {C.}~\bibnamefont {Lei}},
  \bibinfo {author} {\bibfnamefont {A.~H.}\ \bibnamefont {MacDonald}}, \ and\
  \bibinfo {author} {\bibfnamefont {E.~M.}\ \bibnamefont {Hankiewicz}},\
  }\bibfield  {title} {\enquote {\bibinfo {title} {Majorana-based quantum
  computing in nanowire devices},}\ }\href {\doibase
  10.1103/PhysRevB.102.125407} {\bibfield  {journal} {\bibinfo  {journal}
  {Phys. Rev. B}\ }\textbf {\bibinfo {volume} {102}},\ \bibinfo {pages}
  {125407} (\bibinfo {year} {2020})}\BibitemShut {NoStop}%
\bibitem [{\citenamefont {Zilberberg}\ \emph {et~al.}(2008)\citenamefont
  {Zilberberg}, \citenamefont {Braunecker},\ and\ \citenamefont
  {Loss}}]{ZilberbergPRA2008}%
  \BibitemOpen
  \bibfield  {author} {\bibinfo {author} {\bibfnamefont {Oded}\ \bibnamefont
  {Zilberberg}}, \bibinfo {author} {\bibfnamefont {Bernd}\ \bibnamefont
  {Braunecker}}, \ and\ \bibinfo {author} {\bibfnamefont {Daniel}\ \bibnamefont
  {Loss}},\ }\bibfield  {title} {\enquote {\bibinfo {title} {Controlled-{NOT}
  gate for multiparticle qubits and topological quantum computation based on
  parity measurements},}\ }\href {\doibase 10.1103/PhysRevA.77.012327}
  {\bibfield  {journal} {\bibinfo  {journal} {Phys. Rev. A}\ }\textbf {\bibinfo
  {volume} {77}},\ \bibinfo {pages} {012327} (\bibinfo {year}
  {2008})}\BibitemShut {NoStop}%
\bibitem [{\citenamefont {Beenakker}\ \emph {et~al.}(2004)\citenamefont
  {Beenakker}, \citenamefont {DiVincenzo}, \citenamefont {Emary},\ and\
  \citenamefont {Kindermann}}]{BeenakkerPRL2004}%
  \BibitemOpen
  \bibfield  {author} {\bibinfo {author} {\bibfnamefont {C.~W.~J.}\
  \bibnamefont {Beenakker}}, \bibinfo {author} {\bibfnamefont {D.~P.}\
  \bibnamefont {DiVincenzo}}, \bibinfo {author} {\bibfnamefont
  {C.}~\bibnamefont {Emary}}, \ and\ \bibinfo {author} {\bibfnamefont
  {M.}~\bibnamefont {Kindermann}},\ }\bibfield  {title} {\enquote {\bibinfo
  {title} {Charge detection enables free-electron quantum computation},}\
  }\href {\doibase 10.1103/PhysRevLett.93.020501} {\bibfield  {journal}
  {\bibinfo  {journal} {Phys. Rev. Lett.}\ }\textbf {\bibinfo {volume} {93}},\
  \bibinfo {pages} {020501} (\bibinfo {year} {2004})}\BibitemShut {NoStop}%
\bibitem [{\citenamefont {Vijay}\ and\ \citenamefont
  {Fu}(2016)}]{VijayPRB2016}%
  \BibitemOpen
  \bibfield  {author} {\bibinfo {author} {\bibfnamefont {Sagar}\ \bibnamefont
  {Vijay}}\ and\ \bibinfo {author} {\bibfnamefont {Liang}\ \bibnamefont {Fu}},\
  }\bibfield  {title} {\enquote {\bibinfo {title} {Teleportation-based quantum
  information processing with {Majorana} zero modes},}\ }\href {\doibase
  10.1103/PhysRevB.94.235446} {\bibfield  {journal} {\bibinfo  {journal} {Phys.
  Rev. B}\ }\textbf {\bibinfo {volume} {94}},\ \bibinfo {pages} {235446}
  (\bibinfo {year} {2016})}\BibitemShut {NoStop}%
\bibitem [{\citenamefont {Plugge}\ \emph {et~al.}(2017)\citenamefont {Plugge},
  \citenamefont {Rasmussen}, \citenamefont {Egger},\ and\ \citenamefont
  {Flensberg}}]{PluggeNJP2017}%
  \BibitemOpen
  \bibfield  {author} {\bibinfo {author} {\bibfnamefont {Stephan}\ \bibnamefont
  {Plugge}}, \bibinfo {author} {\bibfnamefont {Asbj{\o}rn}\ \bibnamefont
  {Rasmussen}}, \bibinfo {author} {\bibfnamefont {Reinhold}\ \bibnamefont
  {Egger}}, \ and\ \bibinfo {author} {\bibfnamefont {Karsten}\ \bibnamefont
  {Flensberg}},\ }\bibfield  {title} {\enquote {\bibinfo {title} {Majorana box
  qubits},}\ }\href {\doibase 10.1088/1367-2630/aa54e1} {\bibfield  {journal}
  {\bibinfo  {journal} {New Journal of Physics}\ }\textbf {\bibinfo {volume}
  {19}},\ \bibinfo {pages} {012001} (\bibinfo {year} {2017})}\BibitemShut
  {NoStop}%
\bibitem [{\citenamefont {Karzig}\ \emph {et~al.}(2015)\citenamefont {Karzig},
  \citenamefont {Pientka}, \citenamefont {Refael},\ and\ \citenamefont {von
  Oppen}}]{KarzigPRB2015}%
  \BibitemOpen
  \bibfield  {author} {\bibinfo {author} {\bibfnamefont {Torsten}\ \bibnamefont
  {Karzig}}, \bibinfo {author} {\bibfnamefont {Falko}\ \bibnamefont {Pientka}},
  \bibinfo {author} {\bibfnamefont {Gil}\ \bibnamefont {Refael}}, \ and\
  \bibinfo {author} {\bibfnamefont {Felix}\ \bibnamefont {von Oppen}},\
  }\bibfield  {title} {\enquote {\bibinfo {title} {Shortcuts to non-{Abelian}
  braiding},}\ }\href {\doibase 10.1103/PhysRevB.91.201102} {\bibfield
  {journal} {\bibinfo  {journal} {Phys. Rev. B}\ }\textbf {\bibinfo {volume}
  {91}},\ \bibinfo {pages} {201102} (\bibinfo {year} {2015})}\BibitemShut
  {NoStop}%
\bibitem [{\citenamefont {Knapp}\ \emph {et~al.}(2016)\citenamefont {Knapp},
  \citenamefont {Zaletel}, \citenamefont {Liu}, \citenamefont {Cheng},
  \citenamefont {Bonderson},\ and\ \citenamefont {Nayak}}]{KnappPRX2016}%
  \BibitemOpen
  \bibfield  {author} {\bibinfo {author} {\bibfnamefont {Christina}\
  \bibnamefont {Knapp}}, \bibinfo {author} {\bibfnamefont {Michael}\
  \bibnamefont {Zaletel}}, \bibinfo {author} {\bibfnamefont {Dong~E.}\
  \bibnamefont {Liu}}, \bibinfo {author} {\bibfnamefont {Meng}\ \bibnamefont
  {Cheng}}, \bibinfo {author} {\bibfnamefont {Parsa}\ \bibnamefont
  {Bonderson}}, \ and\ \bibinfo {author} {\bibfnamefont {Chetan}\ \bibnamefont
  {Nayak}},\ }\bibfield  {title} {\enquote {\bibinfo {title} {The nature and
  correction of diabatic errors in anyon braiding},}\ }\href {\doibase
  10.1103/PhysRevX.6.041003} {\bibfield  {journal} {\bibinfo  {journal} {Phys.
  Rev. X}\ }\textbf {\bibinfo {volume} {6}},\ \bibinfo {pages} {041003}
  (\bibinfo {year} {2016})}\BibitemShut {NoStop}%
\bibitem [{\citenamefont {Nag}\ and\ \citenamefont {Sau}(2019)}]{NagPRB2019}%
  \BibitemOpen
  \bibfield  {author} {\bibinfo {author} {\bibfnamefont {Amit}\ \bibnamefont
  {Nag}}\ and\ \bibinfo {author} {\bibfnamefont {Jay~D.}\ \bibnamefont {Sau}},\
  }\bibfield  {title} {\enquote {\bibinfo {title} {Diabatic errors in
  {Majorana} braiding with bosonic bath},}\ }\href {\doibase
  10.1103/PhysRevB.100.014511} {\bibfield  {journal} {\bibinfo  {journal}
  {Phys. Rev. B}\ }\textbf {\bibinfo {volume} {100}},\ \bibinfo {pages}
  {014511} (\bibinfo {year} {2019})}\BibitemShut {NoStop}%
\bibitem [{\citenamefont {Stenger}\ \emph {et~al.}(2021)\citenamefont
  {Stenger}, \citenamefont {Bronn}, \citenamefont {Egger},\ and\ \citenamefont
  {Pekker}}]{StengerPRR2021}%
  \BibitemOpen
  \bibfield  {author} {\bibinfo {author} {\bibfnamefont {John P.~T.}\
  \bibnamefont {Stenger}}, \bibinfo {author} {\bibfnamefont {Nicholas~T.}\
  \bibnamefont {Bronn}}, \bibinfo {author} {\bibfnamefont {Daniel~J.}\
  \bibnamefont {Egger}}, \ and\ \bibinfo {author} {\bibfnamefont {David}\
  \bibnamefont {Pekker}},\ }\bibfield  {title} {\enquote {\bibinfo {title}
  {Simulating the dynamics of braiding of {Majorana} zero modes using an {IBM}
  quantum computer},}\ }\href {\doibase 10.1103/PhysRevResearch.3.033171}
  {\bibfield  {journal} {\bibinfo  {journal} {Phys. Rev. Research}\ }\textbf
  {\bibinfo {volume} {3}},\ \bibinfo {pages} {033171} (\bibinfo {year}
  {2021})}\BibitemShut {NoStop}%
\bibitem [{\citenamefont {Amorim}\ \emph {et~al.}(2015)\citenamefont {Amorim},
  \citenamefont {Ebihara}, \citenamefont {Yamakage}, \citenamefont {Tanaka},\
  and\ \citenamefont {Sato}}]{AmorimPRB2015}%
  \BibitemOpen
  \bibfield  {author} {\bibinfo {author} {\bibfnamefont {C\'assio~Sozinho}\
  \bibnamefont {Amorim}}, \bibinfo {author} {\bibfnamefont {Kazuto}\
  \bibnamefont {Ebihara}}, \bibinfo {author} {\bibfnamefont {Ai}~\bibnamefont
  {Yamakage}}, \bibinfo {author} {\bibfnamefont {Yukio}\ \bibnamefont
  {Tanaka}}, \ and\ \bibinfo {author} {\bibfnamefont {Masatoshi}\ \bibnamefont
  {Sato}},\ }\bibfield  {title} {\enquote {\bibinfo {title} {Majorana braiding
  dynamics in nanowires},}\ }\href {\doibase 10.1103/PhysRevB.91.174305}
  {\bibfield  {journal} {\bibinfo  {journal} {Phys. Rev. B}\ }\textbf {\bibinfo
  {volume} {91}},\ \bibinfo {pages} {174305} (\bibinfo {year}
  {2015})}\BibitemShut {NoStop}%
\bibitem [{\citenamefont {Cheng}\ \emph {et~al.}(2016)\citenamefont {Cheng},
  \citenamefont {He},\ and\ \citenamefont {Kou}}]{ChengPLA2016}%
  \BibitemOpen
  \bibfield  {author} {\bibinfo {author} {\bibfnamefont {Qiu-Bo}\ \bibnamefont
  {Cheng}}, \bibinfo {author} {\bibfnamefont {Jing}\ \bibnamefont {He}}, \ and\
  \bibinfo {author} {\bibfnamefont {Su-Peng}\ \bibnamefont {Kou}},\ }\bibfield
  {title} {\enquote {\bibinfo {title} {Verifying non-{Abelian} statistics by
  numerical braiding {Majorana} fermions},}\ }\href {\doibase
  https://doi.org/10.1016/j.physleta.2015.11.030} {\bibfield  {journal}
  {\bibinfo  {journal} {Physics Letters A}\ }\textbf {\bibinfo {volume}
  {380}},\ \bibinfo {pages} {779--782} (\bibinfo {year} {2016})}\BibitemShut
  {NoStop}%
\bibitem [{\citenamefont {Sekania}\ \emph {et~al.}(2017)\citenamefont
  {Sekania}, \citenamefont {Plugge}, \citenamefont {Greiter}, \citenamefont
  {Thomale},\ and\ \citenamefont {Schmitteckert}}]{SekaniaPRB2017}%
  \BibitemOpen
  \bibfield  {author} {\bibinfo {author} {\bibfnamefont {Michael}\ \bibnamefont
  {Sekania}}, \bibinfo {author} {\bibfnamefont {Stephan}\ \bibnamefont
  {Plugge}}, \bibinfo {author} {\bibfnamefont {Martin}\ \bibnamefont
  {Greiter}}, \bibinfo {author} {\bibfnamefont {Ronny}\ \bibnamefont
  {Thomale}}, \ and\ \bibinfo {author} {\bibfnamefont {Peter}\ \bibnamefont
  {Schmitteckert}},\ }\bibfield  {title} {\enquote {\bibinfo {title} {Braiding
  errors in interacting {Majorana} quantum wires},}\ }\href {\doibase
  10.1103/PhysRevB.96.094307} {\bibfield  {journal} {\bibinfo  {journal} {Phys.
  Rev. B}\ }\textbf {\bibinfo {volume} {96}},\ \bibinfo {pages} {094307}
  (\bibinfo {year} {2017})}\BibitemShut {NoStop}%
\bibitem [{\citenamefont {Harper}\ \emph {et~al.}(2019)\citenamefont {Harper},
  \citenamefont {Pushp},\ and\ \citenamefont {Roy}}]{HarperPRR2019}%
  \BibitemOpen
  \bibfield  {author} {\bibinfo {author} {\bibfnamefont {Fenner}\ \bibnamefont
  {Harper}}, \bibinfo {author} {\bibfnamefont {Aakash}\ \bibnamefont {Pushp}},
  \ and\ \bibinfo {author} {\bibfnamefont {Rahul}\ \bibnamefont {Roy}},\
  }\bibfield  {title} {\enquote {\bibinfo {title} {Majorana braiding in
  realistic nanowire {Y}-junctions and tuning forks},}\ }\href {\doibase
  10.1103/PhysRevResearch.1.033207} {\bibfield  {journal} {\bibinfo  {journal}
  {Phys. Rev. Research}\ }\textbf {\bibinfo {volume} {1}},\ \bibinfo {pages}
  {033207} (\bibinfo {year} {2019})}\BibitemShut {NoStop}%
\bibitem [{\citenamefont {Leijnse}\ and\ \citenamefont
  {Flensberg}(2012)}]{Leijnse_2012}%
  \BibitemOpen
  \bibfield  {author} {\bibinfo {author} {\bibfnamefont {Martin}\ \bibnamefont
  {Leijnse}}\ and\ \bibinfo {author} {\bibfnamefont {Karsten}\ \bibnamefont
  {Flensberg}},\ }\bibfield  {title} {\enquote {\bibinfo {title} {Introduction
  to topological superconductivity and {Majorana} fermions},}\ }\href {\doibase
  10.1088/0268-1242/27/12/124003} {\bibfield  {journal} {\bibinfo  {journal}
  {Semiconductor Science and Technology}\ }\textbf {\bibinfo {volume} {27}},\
  \bibinfo {pages} {124003} (\bibinfo {year} {2012})}\BibitemShut {NoStop}%
\bibitem [{\citenamefont {Sau}\ and\ \citenamefont
  {Sarma}(2012)}]{SauNatComm2012}%
  \BibitemOpen
  \bibfield  {author} {\bibinfo {author} {\bibfnamefont {Jay~D.}\ \bibnamefont
  {Sau}}\ and\ \bibinfo {author} {\bibfnamefont {S.~Das}\ \bibnamefont
  {Sarma}},\ }\bibfield  {title} {\enquote {\bibinfo {title} {Realizing a
  robust practical {Majorana} chain in a quantum-dot-superconductor linear
  array},}\ }\href {\doibase 10.1038/ncomms1966} {\bibfield  {journal}
  {\bibinfo  {journal} {Nature Communications}\ }\textbf {\bibinfo {volume}
  {3}},\ \bibinfo {pages} {964} (\bibinfo {year} {2012})}\BibitemShut {NoStop}%
\bibitem [{\citenamefont {Fulga}\ \emph
  {et~al.}(2013{\natexlab{b}})\citenamefont {Fulga}, \citenamefont {Haim},
  \citenamefont {Akhmerov},\ and\ \citenamefont {Oreg}}]{Fulga_2013}%
  \BibitemOpen
  \bibfield  {author} {\bibinfo {author} {\bibfnamefont {Ion~C}\ \bibnamefont
  {Fulga}}, \bibinfo {author} {\bibfnamefont {Arbel}\ \bibnamefont {Haim}},
  \bibinfo {author} {\bibfnamefont {Anton~R}\ \bibnamefont {Akhmerov}}, \ and\
  \bibinfo {author} {\bibfnamefont {Yuval}\ \bibnamefont {Oreg}},\ }\bibfield
  {title} {\enquote {\bibinfo {title} {Adaptive tuning of {Majorana} fermions
  in a quantum dot chain},}\ }\href {\doibase 10.1088/1367-2630/15/4/045020}
  {\bibfield  {journal} {\bibinfo  {journal} {New Journal of Physics}\ }\textbf
  {\bibinfo {volume} {15}},\ \bibinfo {pages} {045020} (\bibinfo {year}
  {2013}{\natexlab{b}})}\BibitemShut {NoStop}%
\bibitem [{\citenamefont {Tsintzis}\ \emph
  {et~al.}(2022{\natexlab{a}})\citenamefont {Tsintzis}, \citenamefont {Souto},\
  and\ \citenamefont {Leijnse}}]{TsintzisPRB2022}%
  \BibitemOpen
  \bibfield  {author} {\bibinfo {author} {\bibfnamefont {Athanasios}\
  \bibnamefont {Tsintzis}}, \bibinfo {author} {\bibfnamefont {Rub\'en~Seoane}\
  \bibnamefont {Souto}}, \ and\ \bibinfo {author} {\bibfnamefont {Martin}\
  \bibnamefont {Leijnse}},\ }\bibfield  {title} {\enquote {\bibinfo {title}
  {{Creating and detecting poor man's Majorana bound states in interacting
  quantum dots}},}\ }\href {\doibase 10.1103/PhysRevB.106.L201404} {\bibfield
  {journal} {\bibinfo  {journal} {Phys. Rev. B}\ }\textbf {\bibinfo {volume}
  {106}},\ \bibinfo {pages} {L201404} (\bibinfo {year}
  {2022}{\natexlab{a}})}\BibitemShut {NoStop}%
\bibitem [{\citenamefont {Liu}\ \emph {et~al.}(2022)\citenamefont {Liu},
  \citenamefont {Wang}, \citenamefont {Dvir},\ and\ \citenamefont
  {Wimmer}}]{ChunXiaoLiuPRL2022}%
  \BibitemOpen
  \bibfield  {author} {\bibinfo {author} {\bibfnamefont {Chun-Xiao}\
  \bibnamefont {Liu}}, \bibinfo {author} {\bibfnamefont {Guanzhong}\
  \bibnamefont {Wang}}, \bibinfo {author} {\bibfnamefont {Tom}\ \bibnamefont
  {Dvir}}, \ and\ \bibinfo {author} {\bibfnamefont {Michael}\ \bibnamefont
  {Wimmer}},\ }\bibfield  {title} {\enquote {\bibinfo {title} {{Tunable
  Superconducting Coupling of Quantum Dots via Andreev Bound States in
  Semiconductor-Superconductor Nanowires}},}\ }\href {\doibase
  10.1103/PhysRevLett.129.267701} {\bibfield  {journal} {\bibinfo  {journal}
  {Phys. Rev. Lett.}\ }\textbf {\bibinfo {volume} {129}},\ \bibinfo {pages}
  {267701} (\bibinfo {year} {2022})}\BibitemShut {NoStop}%
\bibitem [{\citenamefont {Mills}\ \emph {et~al.}(2019)\citenamefont {Mills},
  \citenamefont {Zajac}, \citenamefont {Gullans}, \citenamefont {Schupp},
  \citenamefont {Hazard},\ and\ \citenamefont {Petta}}]{MillsNatComm2019}%
  \BibitemOpen
  \bibfield  {author} {\bibinfo {author} {\bibfnamefont {A.~R.}\ \bibnamefont
  {Mills}}, \bibinfo {author} {\bibfnamefont {D.~M.}\ \bibnamefont {Zajac}},
  \bibinfo {author} {\bibfnamefont {M.~J.}\ \bibnamefont {Gullans}}, \bibinfo
  {author} {\bibfnamefont {F.~J.}\ \bibnamefont {Schupp}}, \bibinfo {author}
  {\bibfnamefont {T.~M.}\ \bibnamefont {Hazard}}, \ and\ \bibinfo {author}
  {\bibfnamefont {J.~R.}\ \bibnamefont {Petta}},\ }\bibfield  {title} {\enquote
  {\bibinfo {title} {Shuttling a single charge across a one-dimensional array
  of silicon quantum dots},}\ }\href {\doibase 10.1038/s41467-019-08970-z}
  {\bibfield  {journal} {\bibinfo  {journal} {Nature Communications}\ }\textbf
  {\bibinfo {volume} {10}},\ \bibinfo {pages} {1063} (\bibinfo {year}
  {2019})}\BibitemShut {NoStop}%
\bibitem [{\citenamefont {Philips}\ \emph {et~al.}(2022)\citenamefont
  {Philips}, \citenamefont {Mądzik}, \citenamefont {Amitonov}, \citenamefont
  {de~Snoo}, \citenamefont {Russ}, \citenamefont {Kalhor}, \citenamefont
  {Volk}, \citenamefont {Lawrie}, \citenamefont {Brousse}, \citenamefont
  {Tryputen}, \citenamefont {Wuetz}, \citenamefont {Sammak}, \citenamefont
  {Veldhorst}, \citenamefont {Scappucci},\ and\ \citenamefont
  {Vandersypen}}]{PhilipsArxiv2022}%
  \BibitemOpen
  \bibfield  {author} {\bibinfo {author} {\bibfnamefont {Stephan G.~J.}\
  \bibnamefont {Philips}}, \bibinfo {author} {\bibfnamefont {Mateusz~T.}\
  \bibnamefont {Mądzik}}, \bibinfo {author} {\bibfnamefont {Sergey~V.}\
  \bibnamefont {Amitonov}}, \bibinfo {author} {\bibfnamefont {Sander~L.}\
  \bibnamefont {de~Snoo}}, \bibinfo {author} {\bibfnamefont {Maximilian}\
  \bibnamefont {Russ}}, \bibinfo {author} {\bibfnamefont {Nima}\ \bibnamefont
  {Kalhor}}, \bibinfo {author} {\bibfnamefont {Christian}\ \bibnamefont
  {Volk}}, \bibinfo {author} {\bibfnamefont {William I.~L.}\ \bibnamefont
  {Lawrie}}, \bibinfo {author} {\bibfnamefont {Delphine}\ \bibnamefont
  {Brousse}}, \bibinfo {author} {\bibfnamefont {Larysa}\ \bibnamefont
  {Tryputen}}, \bibinfo {author} {\bibfnamefont {Brian~Paquelet}\ \bibnamefont
  {Wuetz}}, \bibinfo {author} {\bibfnamefont {Amir}\ \bibnamefont {Sammak}},
  \bibinfo {author} {\bibfnamefont {Menno}\ \bibnamefont {Veldhorst}}, \bibinfo
  {author} {\bibfnamefont {Giordano}\ \bibnamefont {Scappucci}}, \ and\
  \bibinfo {author} {\bibfnamefont {Lieven M.~K.}\ \bibnamefont
  {Vandersypen}},\ }\href {\doibase 10.48550/ARXIV.2202.09252} {\enquote
  {\bibinfo {title} {Universal control of a six-qubit quantum processor in
  silicon},}\ } (\bibinfo {year} {2022})\BibitemShut {NoStop}%
\bibitem [{\citenamefont {Weinstein}\ \emph {et~al.}(2023)\citenamefont
  {Weinstein}, \citenamefont {Reed}, \citenamefont {Jones}, \citenamefont
  {Andrews}, \citenamefont {Barnes}, \citenamefont {Blumoff}, \citenamefont
  {Euliss}, \citenamefont {Eng}, \citenamefont {Fong}, \citenamefont {Ha},
  \citenamefont {Hulbert}, \citenamefont {Jackson}, \citenamefont {Jura},
  \citenamefont {Keating}, \citenamefont {Kerckhoff}, \citenamefont {Kiselev},
  \citenamefont {Matten}, \citenamefont {Sabbir}, \citenamefont {Smith},
  \citenamefont {Wright}, \citenamefont {Rakher}, \citenamefont {Ladd},\ and\
  \citenamefont {Borselli}}]{Weinstein}%
  \BibitemOpen
  \bibfield  {author} {\bibinfo {author} {\bibfnamefont {Aaron~J.}\
  \bibnamefont {Weinstein}}, \bibinfo {author} {\bibfnamefont {Matthew~D.}\
  \bibnamefont {Reed}}, \bibinfo {author} {\bibfnamefont {Aaron~M.}\
  \bibnamefont {Jones}}, \bibinfo {author} {\bibfnamefont {Reed~W.}\
  \bibnamefont {Andrews}}, \bibinfo {author} {\bibfnamefont {David}\
  \bibnamefont {Barnes}}, \bibinfo {author} {\bibfnamefont {Jacob~Z.}\
  \bibnamefont {Blumoff}}, \bibinfo {author} {\bibfnamefont {Larken~E.}\
  \bibnamefont {Euliss}}, \bibinfo {author} {\bibfnamefont {Kevin}\
  \bibnamefont {Eng}}, \bibinfo {author} {\bibfnamefont {Bryan~H.}\
  \bibnamefont {Fong}}, \bibinfo {author} {\bibfnamefont {Sieu~D.}\
  \bibnamefont {Ha}}, \bibinfo {author} {\bibfnamefont {Daniel~R.}\
  \bibnamefont {Hulbert}}, \bibinfo {author} {\bibfnamefont {Clayton A.~C.}\
  \bibnamefont {Jackson}}, \bibinfo {author} {\bibfnamefont {Michael}\
  \bibnamefont {Jura}}, \bibinfo {author} {\bibfnamefont {Tyler~E.}\
  \bibnamefont {Keating}}, \bibinfo {author} {\bibfnamefont {Joseph}\
  \bibnamefont {Kerckhoff}}, \bibinfo {author} {\bibfnamefont {Andrey~A.}\
  \bibnamefont {Kiselev}}, \bibinfo {author} {\bibfnamefont {Justine}\
  \bibnamefont {Matten}}, \bibinfo {author} {\bibfnamefont {Golam}\
  \bibnamefont {Sabbir}}, \bibinfo {author} {\bibfnamefont {Aaron}\
  \bibnamefont {Smith}}, \bibinfo {author} {\bibfnamefont {Jeffrey}\
  \bibnamefont {Wright}}, \bibinfo {author} {\bibfnamefont {Matthew~T.}\
  \bibnamefont {Rakher}}, \bibinfo {author} {\bibfnamefont {Thaddeus~D.}\
  \bibnamefont {Ladd}}, \ and\ \bibinfo {author} {\bibfnamefont {Matthew~G.}\
  \bibnamefont {Borselli}},\ }\bibfield  {title} {\enquote {\bibinfo {title}
  {Universal logic with encoded spin qubits in silicon},}\ }\href {\doibase
  10.1038/s41586-023-05777-3} {\bibfield  {journal} {\bibinfo  {journal}
  {Nature}\ }\textbf {\bibinfo {volume} {615}},\ \bibinfo {pages} {817--822}
  (\bibinfo {year} {2023})}\BibitemShut {NoStop}%
\bibitem [{\citenamefont {Wu}\ \emph {et~al.}(2021)\citenamefont {Wu},
  \citenamefont {Zhang}, \citenamefont {Stenger}, \citenamefont {Su},
  \citenamefont {Chen}, \citenamefont {Badawy}, \citenamefont {Gazibegovic},
  \citenamefont {Bakkers},\ and\ \citenamefont {Frolov}}]{WuArxiv2021}%
  \BibitemOpen
  \bibfield  {author} {\bibinfo {author} {\bibfnamefont {Hao}\ \bibnamefont
  {Wu}}, \bibinfo {author} {\bibfnamefont {Po}~\bibnamefont {Zhang}}, \bibinfo
  {author} {\bibfnamefont {John P.~T.}\ \bibnamefont {Stenger}}, \bibinfo
  {author} {\bibfnamefont {Zhaoen}\ \bibnamefont {Su}}, \bibinfo {author}
  {\bibfnamefont {Jun}\ \bibnamefont {Chen}}, \bibinfo {author} {\bibfnamefont
  {Ghada}\ \bibnamefont {Badawy}}, \bibinfo {author} {\bibfnamefont {Sasa}\
  \bibnamefont {Gazibegovic}}, \bibinfo {author} {\bibfnamefont {Erik P.
  A.~M.}\ \bibnamefont {Bakkers}}, \ and\ \bibinfo {author} {\bibfnamefont
  {Sergey~M.}\ \bibnamefont {Frolov}},\ }\href {\doibase
  10.48550/ARXIV.2105.08636} {\enquote {\bibinfo {title} {Triple {Andreev} dot
  chains in semiconductor nanowires},}\ } (\bibinfo {year} {2021})\BibitemShut
  {NoStop}%
\bibitem [{\citenamefont {Dvir}\ \emph {et~al.}(2023)\citenamefont {Dvir},
  \citenamefont {Wang}, \citenamefont {van Loo}, \citenamefont {Liu},
  \citenamefont {Mazur}, \citenamefont {Bordin}, \citenamefont {ten Haaf},
  \citenamefont {Wang}, \citenamefont {van Driel}, \citenamefont {Zatelli},
  \citenamefont {Li}, \citenamefont {Malinowski}, \citenamefont {Gazibegovic},
  \citenamefont {Badawy}, \citenamefont {Bakkers}, \citenamefont {Wimmer},\
  and\ \citenamefont {Kouwenhoven}}]{Dvir}%
  \BibitemOpen
  \bibfield  {author} {\bibinfo {author} {\bibfnamefont {Tom}\ \bibnamefont
  {Dvir}}, \bibinfo {author} {\bibfnamefont {Guanzhong}\ \bibnamefont {Wang}},
  \bibinfo {author} {\bibfnamefont {Nick}\ \bibnamefont {van Loo}}, \bibinfo
  {author} {\bibfnamefont {Chun-Xiao}\ \bibnamefont {Liu}}, \bibinfo {author}
  {\bibfnamefont {Grzegorz~P.}\ \bibnamefont {Mazur}}, \bibinfo {author}
  {\bibfnamefont {Alberto}\ \bibnamefont {Bordin}}, \bibinfo {author}
  {\bibfnamefont {Sebastiaan L.~D.}\ \bibnamefont {ten Haaf}}, \bibinfo
  {author} {\bibfnamefont {Ji-Yin}\ \bibnamefont {Wang}}, \bibinfo {author}
  {\bibfnamefont {David}\ \bibnamefont {van Driel}}, \bibinfo {author}
  {\bibfnamefont {Francesco}\ \bibnamefont {Zatelli}}, \bibinfo {author}
  {\bibfnamefont {Xiang}\ \bibnamefont {Li}}, \bibinfo {author} {\bibfnamefont
  {Filip~K.}\ \bibnamefont {Malinowski}}, \bibinfo {author} {\bibfnamefont
  {Sasa}\ \bibnamefont {Gazibegovic}}, \bibinfo {author} {\bibfnamefont
  {Ghada}\ \bibnamefont {Badawy}}, \bibinfo {author} {\bibfnamefont {Erik P.
  A.~M.}\ \bibnamefont {Bakkers}}, \bibinfo {author} {\bibfnamefont {Michael}\
  \bibnamefont {Wimmer}}, \ and\ \bibinfo {author} {\bibfnamefont {Leo~P.}\
  \bibnamefont {Kouwenhoven}},\ }\bibfield  {title} {\enquote {\bibinfo {title}
  {Realization of a minimal {Kitaev} chain in coupled quantum dots},}\ }\href
  {\doibase 10.1038/s41586-022-05585-1} {\bibfield  {journal} {\bibinfo
  {journal} {Nature}\ }\textbf {\bibinfo {volume} {614}},\ \bibinfo {pages}
  {445--450} (\bibinfo {year} {2023})}\BibitemShut {NoStop}%
\bibitem [{\citenamefont {Cywi\ifmmode~\acute{n}\else \'{n}\fi{}ski}\ \emph
  {et~al.}(2008)\citenamefont {Cywi\ifmmode~\acute{n}\else \'{n}\fi{}ski},
  \citenamefont {Lutchyn}, \citenamefont {Nave},\ and\ \citenamefont
  {Das~Sarma}}]{CywinskiPRB2008}%
  \BibitemOpen
  \bibfield  {author} {\bibinfo {author} {\bibfnamefont {\L{}ukasz}\
  \bibnamefont {Cywi\ifmmode~\acute{n}\else \'{n}\fi{}ski}}, \bibinfo {author}
  {\bibfnamefont {Roman~M.}\ \bibnamefont {Lutchyn}}, \bibinfo {author}
  {\bibfnamefont {Cody~P.}\ \bibnamefont {Nave}}, \ and\ \bibinfo {author}
  {\bibfnamefont {S.}~\bibnamefont {Das~Sarma}},\ }\bibfield  {title} {\enquote
  {\bibinfo {title} {How to enhance dephasing time in superconducting
  qubits},}\ }\href {\doibase 10.1103/PhysRevB.77.174509} {\bibfield  {journal}
  {\bibinfo  {journal} {Phys. Rev. B}\ }\textbf {\bibinfo {volume} {77}},\
  \bibinfo {pages} {174509} (\bibinfo {year} {2008})}\BibitemShut {NoStop}%
\bibitem [{\citenamefont {Dial}\ \emph {et~al.}(2013)\citenamefont {Dial},
  \citenamefont {Shulman}, \citenamefont {Harvey}, \citenamefont {Bluhm},
  \citenamefont {Umansky},\ and\ \citenamefont {Yacoby}}]{DialPRL2013}%
  \BibitemOpen
  \bibfield  {author} {\bibinfo {author} {\bibfnamefont {O.~E.}\ \bibnamefont
  {Dial}}, \bibinfo {author} {\bibfnamefont {M.~D.}\ \bibnamefont {Shulman}},
  \bibinfo {author} {\bibfnamefont {S.~P.}\ \bibnamefont {Harvey}}, \bibinfo
  {author} {\bibfnamefont {H.}~\bibnamefont {Bluhm}}, \bibinfo {author}
  {\bibfnamefont {V.}~\bibnamefont {Umansky}}, \ and\ \bibinfo {author}
  {\bibfnamefont {A.}~\bibnamefont {Yacoby}},\ }\bibfield  {title} {\enquote
  {\bibinfo {title} {Charge noise spectroscopy using coherent exchange
  oscillations in a singlet-triplet qubit},}\ }\href {\doibase
  10.1103/PhysRevLett.110.146804} {\bibfield  {journal} {\bibinfo  {journal}
  {Phys. Rev. Lett.}\ }\textbf {\bibinfo {volume} {110}},\ \bibinfo {pages}
  {146804} (\bibinfo {year} {2013})}\BibitemShut {NoStop}%
\bibitem [{\citenamefont {Yoneda}\ \emph {et~al.}(2018)\citenamefont {Yoneda},
  \citenamefont {Takeda}, \citenamefont {Otsuka}, \citenamefont {Nakajima},
  \citenamefont {Delbecq}, \citenamefont {Allison}, \citenamefont {Honda},
  \citenamefont {Kodera}, \citenamefont {Oda}, \citenamefont {Hoshi},
  \citenamefont {Usami}, \citenamefont {Itoh},\ and\ \citenamefont
  {Tarucha}}]{YonedaNatNano2018}%
  \BibitemOpen
  \bibfield  {author} {\bibinfo {author} {\bibfnamefont {Jun}\ \bibnamefont
  {Yoneda}}, \bibinfo {author} {\bibfnamefont {Kenta}\ \bibnamefont {Takeda}},
  \bibinfo {author} {\bibfnamefont {Tomohiro}\ \bibnamefont {Otsuka}}, \bibinfo
  {author} {\bibfnamefont {Takashi}\ \bibnamefont {Nakajima}}, \bibinfo
  {author} {\bibfnamefont {Matthieu~R.}\ \bibnamefont {Delbecq}}, \bibinfo
  {author} {\bibfnamefont {Giles}\ \bibnamefont {Allison}}, \bibinfo {author}
  {\bibfnamefont {Takumu}\ \bibnamefont {Honda}}, \bibinfo {author}
  {\bibfnamefont {Tetsuo}\ \bibnamefont {Kodera}}, \bibinfo {author}
  {\bibfnamefont {Shunri}\ \bibnamefont {Oda}}, \bibinfo {author}
  {\bibfnamefont {Yusuke}\ \bibnamefont {Hoshi}}, \bibinfo {author}
  {\bibfnamefont {Noritaka}\ \bibnamefont {Usami}}, \bibinfo {author}
  {\bibfnamefont {Kohei~M.}\ \bibnamefont {Itoh}}, \ and\ \bibinfo {author}
  {\bibfnamefont {Seigo}\ \bibnamefont {Tarucha}},\ }\bibfield  {title}
  {\enquote {\bibinfo {title} {A quantum-dot spin qubit with coherence limited
  by charge noise and fidelity higher than 99.9{\%}},}\ }\href {\doibase
  10.1038/s41565-017-0014-x} {\bibfield  {journal} {\bibinfo  {journal} {Nature
  Nanotechnology}\ }\textbf {\bibinfo {volume} {13}},\ \bibinfo {pages}
  {102--106} (\bibinfo {year} {2018})}\BibitemShut {NoStop}%
\bibitem [{\citenamefont {Shnirman}\ \emph {et~al.}(2002)\citenamefont
  {Shnirman}, \citenamefont {Makhlin},\ and\ \citenamefont
  {Sch\"on}}]{ShnirmanPhysScri2002}%
  \BibitemOpen
  \bibfield  {author} {\bibinfo {author} {\bibfnamefont {Alexander}\
  \bibnamefont {Shnirman}}, \bibinfo {author} {\bibfnamefont {Yuriy}\
  \bibnamefont {Makhlin}}, \ and\ \bibinfo {author} {\bibfnamefont {Gerd}\
  \bibnamefont {Sch\"on}},\ }\bibfield  {title} {\enquote {\bibinfo {title}
  {Noise and decoherence in quantum two-level systems},}\ }\href {\doibase
  10.1238/physica.topical.102a00147} {\bibfield  {journal} {\bibinfo  {journal}
  {Physica Scripta}\ }\textbf {\bibinfo {volume} {T102}},\ \bibinfo {pages}
  {147} (\bibinfo {year} {2002})}\BibitemShut {NoStop}%
\bibitem [{\citenamefont {Freeman}\ \emph {et~al.}(2016)\citenamefont
  {Freeman}, \citenamefont {Schoenfield},\ and\ \citenamefont
  {Jiang}}]{FreemanAPL2016}%
  \BibitemOpen
  \bibfield  {author} {\bibinfo {author} {\bibfnamefont {Blake~M.}\
  \bibnamefont {Freeman}}, \bibinfo {author} {\bibfnamefont {Joshua~S.}\
  \bibnamefont {Schoenfield}}, \ and\ \bibinfo {author} {\bibfnamefont
  {HongWen}\ \bibnamefont {Jiang}},\ }\bibfield  {title} {\enquote {\bibinfo
  {title} {Comparison of low frequency charge noise in identically patterned
  {Si}/{SiO}$_2$ and {Si}/{SiGe} quantum dots},}\ }\href {\doibase
  10.1063/1.4954700} {\bibfield  {journal} {\bibinfo  {journal} {Applied
  Physics Letters}\ }\textbf {\bibinfo {volume} {108}},\ \bibinfo {pages}
  {253108} (\bibinfo {year} {2016})},\ \Eprint
  {http://arxiv.org/abs/https://doi.org/10.1063/1.4954700}
  {https://doi.org/10.1063/1.4954700} \BibitemShut {NoStop}%
\bibitem [{\citenamefont {Het\'enyi}\ \emph {et~al.}(2019)\citenamefont
  {Het\'enyi}, \citenamefont {Boross},\ and\ \citenamefont
  {P\'alyi}}]{HetenyiPRB2019}%
  \BibitemOpen
  \bibfield  {author} {\bibinfo {author} {\bibfnamefont {Bence}\ \bibnamefont
  {Het\'enyi}}, \bibinfo {author} {\bibfnamefont {P\'eter}\ \bibnamefont
  {Boross}}, \ and\ \bibinfo {author} {\bibfnamefont {Andr\'as}\ \bibnamefont
  {P\'alyi}},\ }\bibfield  {title} {\enquote {\bibinfo {title}
  {Hyperfine-assisted decoherence of a phosphorus nuclear-spin qubit in
  silicon},}\ }\href {\doibase 10.1103/PhysRevB.100.115435} {\bibfield
  {journal} {\bibinfo  {journal} {Phys. Rev. B}\ }\textbf {\bibinfo {volume}
  {100}},\ \bibinfo {pages} {115435} (\bibinfo {year} {2019})}\BibitemShut
  {NoStop}%
\bibitem [{\citenamefont {Krzywda}\ and\ \citenamefont
  {Cywi\ifmmode~\acute{n}\else \'{n}\fi{}ski}(2020)}]{CywinskiPRB2020}%
  \BibitemOpen
  \bibfield  {author} {\bibinfo {author} {\bibfnamefont {Jan~A.}\ \bibnamefont
  {Krzywda}}\ and\ \bibinfo {author} {\bibfnamefont {\L{}ukasz}\ \bibnamefont
  {Cywi\ifmmode~\acute{n}\else \'{n}\fi{}ski}},\ }\bibfield  {title} {\enquote
  {\bibinfo {title} {Adiabatic electron charge transfer between two quantum
  dots in presence of $1/f$ noise},}\ }\href {\doibase
  10.1103/PhysRevB.101.035303} {\bibfield  {journal} {\bibinfo  {journal}
  {Phys. Rev. B}\ }\textbf {\bibinfo {volume} {101}},\ \bibinfo {pages}
  {035303} (\bibinfo {year} {2020})}\BibitemShut {NoStop}%
\bibitem [{\citenamefont {Sz\'echenyi}\ and\ \citenamefont
  {P\'alyi}(2020)}]{SzechenyiPRB2020}%
  \BibitemOpen
  \bibfield  {author} {\bibinfo {author} {\bibfnamefont {G\'abor}\ \bibnamefont
  {Sz\'echenyi}}\ and\ \bibinfo {author} {\bibfnamefont {Andr\'as}\
  \bibnamefont {P\'alyi}},\ }\bibfield  {title} {\enquote {\bibinfo {title}
  {Parity-to-charge conversion for readout of topological {Majorana} qubits},}\
  }\href {\doibase 10.1103/PhysRevB.101.235441} {\bibfield  {journal} {\bibinfo
   {journal} {Phys. Rev. B}\ }\textbf {\bibinfo {volume} {101}},\ \bibinfo
  {pages} {235441} (\bibinfo {year} {2020})}\BibitemShut {NoStop}%
\bibitem [{\citenamefont {Gharavi}\ \emph {et~al.}(2016)\citenamefont
  {Gharavi}, \citenamefont {Hoving},\ and\ \citenamefont
  {Baugh}}]{GharaviPRB2016}%
  \BibitemOpen
  \bibfield  {author} {\bibinfo {author} {\bibfnamefont {Kaveh}\ \bibnamefont
  {Gharavi}}, \bibinfo {author} {\bibfnamefont {Darryl}\ \bibnamefont
  {Hoving}}, \ and\ \bibinfo {author} {\bibfnamefont {Jonathan}\ \bibnamefont
  {Baugh}},\ }\bibfield  {title} {\enquote {\bibinfo {title} {Readout of
  {Majorana} parity states using a quantum dot},}\ }\href {\doibase
  10.1103/PhysRevB.94.155417} {\bibfield  {journal} {\bibinfo  {journal} {Phys.
  Rev. B}\ }\textbf {\bibinfo {volume} {94}},\ \bibinfo {pages} {155417}
  (\bibinfo {year} {2016})}\BibitemShut {NoStop}%
\bibitem [{\citenamefont {Flensberg}(2011)}]{FlensbergPRL2011}%
  \BibitemOpen
  \bibfield  {author} {\bibinfo {author} {\bibfnamefont {Karsten}\ \bibnamefont
  {Flensberg}},\ }\bibfield  {title} {\enquote {\bibinfo {title} {Non-{Abelian}
  operations on {Majorana} fermions via single-charge control},}\ }\href
  {\doibase 10.1103/PhysRevLett.106.090503} {\bibfield  {journal} {\bibinfo
  {journal} {Phys. Rev. Lett.}\ }\textbf {\bibinfo {volume} {106}},\ \bibinfo
  {pages} {090503} (\bibinfo {year} {2011})}\BibitemShut {NoStop}%
\bibitem [{\citenamefont {Li}\ \emph {et~al.}(2018)\citenamefont {Li},
  \citenamefont {Coish}, \citenamefont {Hell}, \citenamefont {Flensberg},\ and\
  \citenamefont {Leijnse}}]{TommyLiPRB2018}%
  \BibitemOpen
  \bibfield  {author} {\bibinfo {author} {\bibfnamefont {Tommy}\ \bibnamefont
  {Li}}, \bibinfo {author} {\bibfnamefont {William~A.}\ \bibnamefont {Coish}},
  \bibinfo {author} {\bibfnamefont {Michael}\ \bibnamefont {Hell}}, \bibinfo
  {author} {\bibfnamefont {Karsten}\ \bibnamefont {Flensberg}}, \ and\ \bibinfo
  {author} {\bibfnamefont {Martin}\ \bibnamefont {Leijnse}},\ }\bibfield
  {title} {\enquote {\bibinfo {title} {Four-{Majorana} qubit with charge
  readout: {Dynamics} and decoherence},}\ }\href {\doibase
  10.1103/PhysRevB.98.205403} {\bibfield  {journal} {\bibinfo  {journal} {Phys.
  Rev. B}\ }\textbf {\bibinfo {volume} {98}},\ \bibinfo {pages} {205403}
  (\bibinfo {year} {2018})}\BibitemShut {NoStop}%
\bibitem [{\citenamefont {Steiner}\ and\ \citenamefont {von
  Oppen}(2020)}]{SteinerPRR2020}%
  \BibitemOpen
  \bibfield  {author} {\bibinfo {author} {\bibfnamefont {Jacob~F.}\
  \bibnamefont {Steiner}}\ and\ \bibinfo {author} {\bibfnamefont {Felix}\
  \bibnamefont {von Oppen}},\ }\bibfield  {title} {\enquote {\bibinfo {title}
  {Readout of {Majorana} qubits},}\ }\href {\doibase
  10.1103/PhysRevResearch.2.033255} {\bibfield  {journal} {\bibinfo  {journal}
  {Phys. Rev. Research}\ }\textbf {\bibinfo {volume} {2}},\ \bibinfo {pages}
  {033255} (\bibinfo {year} {2020})}\BibitemShut {NoStop}%
\bibitem [{\citenamefont {Munk}\ \emph {et~al.}(2020)\citenamefont {Munk},
  \citenamefont {Schulenborg}, \citenamefont {Egger},\ and\ \citenamefont
  {Flensberg}}]{MunkPRR2020}%
  \BibitemOpen
  \bibfield  {author} {\bibinfo {author} {\bibfnamefont {Morten I.~K.}\
  \bibnamefont {Munk}}, \bibinfo {author} {\bibfnamefont {Jens}\ \bibnamefont
  {Schulenborg}}, \bibinfo {author} {\bibfnamefont {Reinhold}\ \bibnamefont
  {Egger}}, \ and\ \bibinfo {author} {\bibfnamefont {Karsten}\ \bibnamefont
  {Flensberg}},\ }\bibfield  {title} {\enquote {\bibinfo {title}
  {Parity-to-charge conversion in {Majorana} qubit readout},}\ }\href {\doibase
  10.1103/PhysRevResearch.2.033254} {\bibfield  {journal} {\bibinfo  {journal}
  {Phys. Rev. Research}\ }\textbf {\bibinfo {volume} {2}},\ \bibinfo {pages}
  {033254} (\bibinfo {year} {2020})}\BibitemShut {NoStop}%
\bibitem [{\citenamefont {Khindanov}\ \emph {et~al.}(2021)\citenamefont
  {Khindanov}, \citenamefont {Pikulin},\ and\ \citenamefont
  {Karzig}}]{KhindanovSciPost2021}%
  \BibitemOpen
  \bibfield  {author} {\bibinfo {author} {\bibfnamefont {Aleksei}\ \bibnamefont
  {Khindanov}}, \bibinfo {author} {\bibfnamefont {Dmitry}\ \bibnamefont
  {Pikulin}}, \ and\ \bibinfo {author} {\bibfnamefont {Torsten}\ \bibnamefont
  {Karzig}},\ }\bibfield  {title} {\enquote {\bibinfo {title} {{Visibility of
  noisy quantum dot-based measurements of {Majorana} qubits}},}\ }\href
  {\doibase 10.21468/SciPostPhys.10.6.127} {\bibfield  {journal} {\bibinfo
  {journal} {SciPost Phys.}\ }\textbf {\bibinfo {volume} {10}},\ \bibinfo
  {pages} {127} (\bibinfo {year} {2021})}\BibitemShut {NoStop}%
\bibitem [{\citenamefont {Boross}\ \emph {et~al.}(2019)\citenamefont {Boross},
  \citenamefont {Asb\'oth}, \citenamefont {Sz\'echenyi}, \citenamefont
  {Oroszl\'any},\ and\ \citenamefont {P\'alyi}}]{BorossPRB2019}%
  \BibitemOpen
  \bibfield  {author} {\bibinfo {author} {\bibfnamefont {P\'eter}\ \bibnamefont
  {Boross}}, \bibinfo {author} {\bibfnamefont {J\'anos~K.}\ \bibnamefont
  {Asb\'oth}}, \bibinfo {author} {\bibfnamefont {G\'abor}\ \bibnamefont
  {Sz\'echenyi}}, \bibinfo {author} {\bibfnamefont {L\'aszl\'o}\ \bibnamefont
  {Oroszl\'any}}, \ and\ \bibinfo {author} {\bibfnamefont {Andr\'as}\
  \bibnamefont {P\'alyi}},\ }\bibfield  {title} {\enquote {\bibinfo {title}
  {Poor man's topological quantum gate based on the {Su-Schrieffer-Heeger}
  model},}\ }\href {\doibase 10.1103/PhysRevB.100.045414} {\bibfield  {journal}
  {\bibinfo  {journal} {Phys. Rev. B}\ }\textbf {\bibinfo {volume} {100}},\
  \bibinfo {pages} {045414} (\bibinfo {year} {2019})}\BibitemShut {NoStop}%
\bibitem [{\citenamefont {Hegde}\ and\ \citenamefont
  {Vishveshwara}(2016)}]{HegdePRB2016}%
  \BibitemOpen
  \bibfield  {author} {\bibinfo {author} {\bibfnamefont {Suraj~S.}\
  \bibnamefont {Hegde}}\ and\ \bibinfo {author} {\bibfnamefont {Smitha}\
  \bibnamefont {Vishveshwara}},\ }\bibfield  {title} {\enquote {\bibinfo
  {title} {Majorana wave-function oscillations, fermion parity switches, and
  disorder in {Kitaev} chains},}\ }\href {\doibase 10.1103/PhysRevB.94.115166}
  {\bibfield  {journal} {\bibinfo  {journal} {Phys. Rev. B}\ }\textbf {\bibinfo
  {volume} {94}},\ \bibinfo {pages} {115166} (\bibinfo {year}
  {2016})}\BibitemShut {NoStop}%
\bibitem [{\citenamefont {Scarlino}\ \emph {et~al.}(2021)\citenamefont
  {Scarlino}, \citenamefont {Ungerer}, \citenamefont {van Woerkom},
  \citenamefont {Mancini}, \citenamefont {Stano}, \citenamefont {Muller},
  \citenamefont {Landig}, \citenamefont {Koski}, \citenamefont {Reichl},
  \citenamefont {Wegscheider}, \citenamefont {Ihn}, \citenamefont {Ensslin},\
  and\ \citenamefont {Wallraff}}]{ScarlinoArxiv2021}%
  \BibitemOpen
  \bibfield  {author} {\bibinfo {author} {\bibfnamefont {P.}~\bibnamefont
  {Scarlino}}, \bibinfo {author} {\bibfnamefont {J.~H.}\ \bibnamefont
  {Ungerer}}, \bibinfo {author} {\bibfnamefont {D.~J.}\ \bibnamefont {van
  Woerkom}}, \bibinfo {author} {\bibfnamefont {M.}~\bibnamefont {Mancini}},
  \bibinfo {author} {\bibfnamefont {P.}~\bibnamefont {Stano}}, \bibinfo
  {author} {\bibfnamefont {C.}~\bibnamefont {Muller}}, \bibinfo {author}
  {\bibfnamefont {A.~J.}\ \bibnamefont {Landig}}, \bibinfo {author}
  {\bibfnamefont {J.~V.}\ \bibnamefont {Koski}}, \bibinfo {author}
  {\bibfnamefont {C.}~\bibnamefont {Reichl}}, \bibinfo {author} {\bibfnamefont
  {W.}~\bibnamefont {Wegscheider}}, \bibinfo {author} {\bibfnamefont
  {T.}~\bibnamefont {Ihn}}, \bibinfo {author} {\bibfnamefont {K.}~\bibnamefont
  {Ensslin}}, \ and\ \bibinfo {author} {\bibfnamefont {A.}~\bibnamefont
  {Wallraff}},\ }\href@noop {} {\enquote {\bibinfo {title} {In-situ tuning of
  the electric dipole strength of a double dot charge qubit: Charge noise
  protection and ultra strong coupling},}\ } (\bibinfo {year} {2021}),\ \Eprint
  {http://arxiv.org/abs/2104.03045} {arXiv:2104.03045} \BibitemShut {NoStop}%
\bibitem [{\citenamefont {Tsintzis}\ \emph
  {et~al.}(2022{\natexlab{b}})\citenamefont {Tsintzis}, \citenamefont {Souto},\
  and\ \citenamefont {Leijnse}}]{Tsintzis}%
  \BibitemOpen
  \bibfield  {author} {\bibinfo {author} {\bibfnamefont {Athanasios}\
  \bibnamefont {Tsintzis}}, \bibinfo {author} {\bibfnamefont {Rub\'en~Seoane}\
  \bibnamefont {Souto}}, \ and\ \bibinfo {author} {\bibfnamefont {Martin}\
  \bibnamefont {Leijnse}},\ }\bibfield  {title} {\enquote {\bibinfo {title}
  {Creating and detecting poor man's {Majorana} bound states in interacting
  quantum dots},}\ }\href {\doibase 10.1103/PhysRevB.106.L201404} {\bibfield
  {journal} {\bibinfo  {journal} {Phys. Rev. B}\ }\textbf {\bibinfo {volume}
  {106}},\ \bibinfo {pages} {L201404} (\bibinfo {year}
  {2022}{\natexlab{b}})}\BibitemShut {NoStop}%
\bibitem [{\citenamefont {Tosi}\ \emph {et~al.}(2017)\citenamefont {Tosi},
  \citenamefont {Mohiyaddin}, \citenamefont {Schmitt}, \citenamefont {Tenberg},
  \citenamefont {Rahman}, \citenamefont {Klimeck},\ and\ \citenamefont
  {Morello}}]{TosiNatComm2017}%
  \BibitemOpen
  \bibfield  {author} {\bibinfo {author} {\bibfnamefont {Guilherme}\
  \bibnamefont {Tosi}}, \bibinfo {author} {\bibfnamefont {Fahd~A.}\
  \bibnamefont {Mohiyaddin}}, \bibinfo {author} {\bibfnamefont {Vivien}\
  \bibnamefont {Schmitt}}, \bibinfo {author} {\bibfnamefont {Stefanie}\
  \bibnamefont {Tenberg}}, \bibinfo {author} {\bibfnamefont {Rajib}\
  \bibnamefont {Rahman}}, \bibinfo {author} {\bibfnamefont {Gerhard}\
  \bibnamefont {Klimeck}}, \ and\ \bibinfo {author} {\bibfnamefont {Andrea}\
  \bibnamefont {Morello}},\ }\bibfield  {title} {\enquote {\bibinfo {title}
  {Silicon quantum processor with robust long-distance qubit couplings},}\
  }\href {\doibase 10.1038/s41467-017-00378-x} {\bibfield  {journal} {\bibinfo
  {journal} {Nature Communications}\ }\textbf {\bibinfo {volume} {8}},\
  \bibinfo {pages} {450} (\bibinfo {year} {2017})}\BibitemShut {NoStop}%
\bibitem [{\citenamefont {Boter}\ \emph {et~al.}(2020)\citenamefont {Boter},
  \citenamefont {Xue}, \citenamefont {Kr\"ahenmann}, \citenamefont {Watson},
  \citenamefont {Premakumar}, \citenamefont {Ward}, \citenamefont {Savage},
  \citenamefont {Lagally}, \citenamefont {Friesen}, \citenamefont
  {Coppersmith}, \citenamefont {Eriksson}, \citenamefont {Joynt},\ and\
  \citenamefont {Vandersypen}}]{BoterPRB2020}%
  \BibitemOpen
  \bibfield  {author} {\bibinfo {author} {\bibfnamefont {Jelmer~M.}\
  \bibnamefont {Boter}}, \bibinfo {author} {\bibfnamefont {Xiao}\ \bibnamefont
  {Xue}}, \bibinfo {author} {\bibfnamefont {Tobias}\ \bibnamefont
  {Kr\"ahenmann}}, \bibinfo {author} {\bibfnamefont {Thomas~F.}\ \bibnamefont
  {Watson}}, \bibinfo {author} {\bibfnamefont {Vickram~N.}\ \bibnamefont
  {Premakumar}}, \bibinfo {author} {\bibfnamefont {Daniel~R.}\ \bibnamefont
  {Ward}}, \bibinfo {author} {\bibfnamefont {Donald~E.}\ \bibnamefont
  {Savage}}, \bibinfo {author} {\bibfnamefont {Max~G.}\ \bibnamefont
  {Lagally}}, \bibinfo {author} {\bibfnamefont {Mark}\ \bibnamefont {Friesen}},
  \bibinfo {author} {\bibfnamefont {Susan~N.}\ \bibnamefont {Coppersmith}},
  \bibinfo {author} {\bibfnamefont {Mark~A.}\ \bibnamefont {Eriksson}},
  \bibinfo {author} {\bibfnamefont {Robert}\ \bibnamefont {Joynt}}, \ and\
  \bibinfo {author} {\bibfnamefont {Lieven M.~K.}\ \bibnamefont
  {Vandersypen}},\ }\bibfield  {title} {\enquote {\bibinfo {title} {Spatial
  noise correlations in a {Si/SiGe} two-qubit device from {Bell} state
  coherences},}\ }\href {\doibase 10.1103/PhysRevB.101.235133} {\bibfield
  {journal} {\bibinfo  {journal} {Phys. Rev. B}\ }\textbf {\bibinfo {volume}
  {101}},\ \bibinfo {pages} {235133} (\bibinfo {year} {2020})}\BibitemShut
  {NoStop}%
\bibitem [{\citenamefont {Makhlin}\ \emph {et~al.}(2003)\citenamefont
  {Makhlin}, \citenamefont {Schon},\ and\ \citenamefont
  {Shnirman}}]{MakhlinNewDirections2003}%
  \BibitemOpen
  \bibfield  {author} {\bibinfo {author} {\bibfnamefont {Yuriy}\ \bibnamefont
  {Makhlin}}, \bibinfo {author} {\bibfnamefont {Gerd}\ \bibnamefont {Schon}}, \
  and\ \bibinfo {author} {\bibfnamefont {Alexander}\ \bibnamefont {Shnirman}},\
  }\bibfield  {title} {\enquote {\bibinfo {title} {Dissipation in {Josephson}
  qubits},}\ }in\ \href@noop {} {\emph {\bibinfo {booktitle} {New Directions in
  Mesoscopic Physics (Towards Nanoscience)}}},\ \bibinfo {editor} {edited by\
  \bibinfo {editor} {\bibfnamefont {R.}~\bibnamefont {Fazio}}, \bibinfo
  {editor} {\bibfnamefont {V.~F.}\ \bibnamefont {Gantmakher}}, \ and\ \bibinfo
  {editor} {\bibfnamefont {Y.}~\bibnamefont {Imry}}}\ (\bibinfo  {publisher}
  {Springer Netherlands},\ \bibinfo {address} {Dordrecht},\ \bibinfo {year}
  {2003})\ pp.\ \bibinfo {pages} {197--224}\BibitemShut {NoStop}%
\bibitem [{\citenamefont {Huang}\ \emph {et~al.}(2019)\citenamefont {Huang},
  \citenamefont {Yang}, \citenamefont {Chen}, \citenamefont {Dzurak},\ and\
  \citenamefont {Goan}}]{HuangPRA2019}%
  \BibitemOpen
  \bibfield  {author} {\bibinfo {author} {\bibfnamefont {Chia-Hsien}\
  \bibnamefont {Huang}}, \bibinfo {author} {\bibfnamefont {Chih-Hwan}\
  \bibnamefont {Yang}}, \bibinfo {author} {\bibfnamefont {Chien-Chang}\
  \bibnamefont {Chen}}, \bibinfo {author} {\bibfnamefont {Andrew~S.}\
  \bibnamefont {Dzurak}}, \ and\ \bibinfo {author} {\bibfnamefont {Hsi-Sheng}\
  \bibnamefont {Goan}},\ }\bibfield  {title} {\enquote {\bibinfo {title}
  {High-fidelity and robust two-qubit gates for quantum-dot spin qubits in
  silicon},}\ }\href {\doibase 10.1103/PhysRevA.99.042310} {\bibfield
  {journal} {\bibinfo  {journal} {Phys. Rev. A}\ }\textbf {\bibinfo {volume}
  {99}},\ \bibinfo {pages} {042310} (\bibinfo {year} {2019})}\BibitemShut
  {NoStop}%
\end{thebibliography}%

\end{document}